\documentclass[a4paper,12pt]{article}
\usepackage{amssymb}
\usepackage{amsmath}
\usepackage{mathtools}
\usepackage{slashed}
\usepackage{color}
\usepackage{float}
\usepackage{indentfirst}
\usepackage{graphicx}
\usepackage{csquotes} 
\usepackage{caption}
\usepackage{comment}
\usepackage{cite}
\usepackage{diagbox}

\renewcommand{\thefootnote}{\fnsymbol{footnote}}

\begin{document}

\title{
\begin{flushright}
\begin{minipage}{0.2\linewidth}
\normalsize
EPHOU-20-012\\
KEK-TH-2275 \\
WU-HEP-20-11\\*[50pt]
\end{minipage}
\end{flushright}
{\Large \bf 
Landscape of Modular Symmetric Flavor Models
\\*[20pt]}}

\author{Keiya Ishiguro$^{a}$\footnote{
E-mail address: keyspire@ruri.waseda.jp
},\,
Tatsuo Kobayashi$^{b}$\footnote{
E-mail address: kobayashi@particle.sci.hokudai.ac.jp
}
\ and\
Hajime~Otsuka$^{c}$\footnote{
E-mail address: hotsuka@post.kek.jp
}\\*[20pt]
$^a${\it \normalsize 
Department of Physics, Waseda University, Tokyo 169-8555, Japan} \\
$^b${\it \normalsize 
Department of Physics, Hokkaido University, Sapporo 060-0810, Japan} \\
$^c${\it \normalsize 
KEK Theory Center, Institute of Particle and Nuclear Studies, KEK,}\\
{\it \normalsize 1-1 Oho, Tsukuba, Ibaraki 305-0801, Japan}}
\maketitle

\date{
\centerline{\small \bf Abstract}
\begin{minipage}{0.9\linewidth}
\medskip 
\medskip 
\small
We study the moduli stabilization from the viewpoint of 
modular flavor symmetries.
We systematically analyze stabilized moduli values in possible configurations 
of flux compactifications, investigating probabilities of moduli values and showing 
which moduli values are favorable from our moduli stabilization.
Then, we 
examine their implications on modular symmetric flavor models.
It is found that distributions of complex structure modulus $\tau$ determining the flavor structure are 
clustered at a fixed point with the residual $\mathbb{Z}_3$ symmetry in the $SL(2,\mathbb{Z})$ fundamental region.
Also, they are clustered at other specific points such as intersecting points between $|\tau|^2=k/2$ and 
${\rm Re}\,\tau=0,\pm 1/4, \pm1/2$, although their probabilities are less than the $\mathbb{Z}_3$ fixed point.
In general, CP-breaking vacua in the complex structure modulus are statistically disfavored in the string landscape. 
Among CP-breaking vacua, the values ${\rm Re}\,\tau=\pm 1/4$ are most favorable in particular when 
the axio-dilaton $S$ is stabilized at ${\rm Re}\,S=\pm 1/4$.
That shows a strong correlation between CP phases originated from string moduli.
\end{minipage}
}

\renewcommand{\thefootnote}{\arabic{footnote}}
\setcounter{footnote}{0}
\thispagestyle{empty}
\clearpage
\addtocounter{page}{-1}

\tableofcontents

\section{Introduction}
\label{sec:1}

In particle physics, it is one of important issues to understand the origin of 
the flavor structure in the quark and lepton sectors, that is, 
the hierarchies of quark and lepton masses, their mixing angles and CP phases.
Indeed, various scenarios have been proposed and studied.
Among them, the approach by non-Abelian discrete flavor symmetries is 
one of interesting approaches, and many studies have been carried out 
by assuming various non-Abelian discrete flavor symmetries such as 
$A_N, S_N, \Delta(6N^2)$ \cite{
	Altarelli:2010gt,Ishimori:2010au,Hernandez:2012ra,
	King:2013eh,King:2014nza}.
For example, the $A_4$ flavor symmetry is one of most extensively studied models.
It is a symmetry of tetrahedron.
Thus, geometrical symmetries can be origins of these non-Abelian discrete flavor symmetries 
in higher-dimensional theories such as superstring theory \cite{Kobayashi:2006wq,Abe:2009vi}.
Symmetries are important tools to connect physics between low-energy and high-energy scales.

The torus and orbifold compactifications also have another  geometrical symmetry called  
the modular symmetry, which  corresponds to a basis change of cycles on the torus.
The two-dimensional (2D) tours has the modular symmetry $SL(2,\mathbb{Z})$, and higher-dimensional tori have larger modular symmetries, although the factorizable torus, e.g. $T^2\times T^2 \times T^2$, has the product symmetry of $SL(2,\mathbb{Z})$.
Furthermore, the symplectic modular symmetry appears in resolutions of 
toroidal orbifolds such as Calabi-Yau compactifications~\cite{Strominger:1990pd,Candelas:1990pi}.

The modular group also transforms non-trivially zero-modes corresponding to quarks and leptons in 
four-dimensional (4D) effective field theory.
That is the flavor symmetry.
Yukawa couplings as well as higher-order couplings are functions of moduli.
Thus, these couplings also transform non-trivially under the modular flavor symmetry.
So far, the modular flavor symmetry are discussed in the context 
of Narain lattice in the heterotic string theory \cite{Baur:2019kwi,Baur:2019iai,Nilles:2020tdp,Nilles:2020gvu} and magnetized D-brane 
models in Type IIB string theory~\cite{Kobayashi:2018rad,Ohki:2020bpo,Kikuchi:2020frp,Kikuchi:2020nxn} from the top-down approach. 
Furthermore, unification of the modular symmetry and the 4D CP is also developed in the heterotic string theory and in Type IIB string theory, 
where the 4D CP is identified with an outer automorphism of the 
$SL(2,\mathbb{Z})$ modular group on toroidal background \cite{Nilles:2018wex,Novichkov:2019sqv} and $Sp(2n,\mathbb{Z})$ 
modular group on Calabi-Yau threefolds~\cite{Ishiguro:2020nuf}. 
That can be an origin of generalized CP symmetry \cite{Feruglio:2012cw,Holthausen:2012dk,Chen:2014tpa,Novichkov:2019sqv}.

Interestingly, the modular group has finite subgroups such as $S_3, A_4, S_4, A_5, \Delta(96), \Delta(384)$ \cite{deAdelhartToorop:2011re}.
These discrete groups have been used to construct phenomenologically interesting flavor models 
in the bottom-up approach \cite{
	Altarelli:2010gt,Ishimori:2010au,Hernandez:2012ra,
	King:2013eh,King:2014nza}.
Motivated by the above aspects, a modular symmetric flavor model was proposed in Ref.~\cite{Feruglio:2017spp} 
as a phenomenological model building in the bottom-up approach, and 
recently modular flavor symmetries are studied extensively.
(See for early works \cite{Kobayashi:2018vbk,Novichkov:2018ovf}.)
Also such studies are extended to covering groups of $\Gamma_N$ \cite{Liu:2019khw}.
In contrast with well-studied models with discrete flavor symmetries, 
these modular symmetric flavor models are controlled 
by the modular symmetry, and 
Yukawa couplings as well as higher-order couplings are written by modular forms, which are functions of 
the modulus $\tau$.
A vacuum expectation value of the modulus $\tau$ induces the 
spontaneously symmetry breaking of the modular symmetry, and consequently 
determines the mass hierarchy and mixing angles of quarks and leptons. 
Also, the axionic part of $\tau$ determines CP phases.
Hence, the value of $\tau$ is the keypoint to realize realistic fermion mass matrices in these modular flavor models. 
In the bottom-up approach, the whole fundamental region of $SL(2,\mathbb{Z})$
is often searched in order to compare the predictions of modular flavor models with the observational data.
There is no guideline to pick up particular points in the moduli space. 
Furthermore, it is unclear whether the phenomenologically favored moduli values discussed in the bottom-up approach 
can be realized by some mechanism.
In this respect, the ultra-violet theory would provide us with a hint on this problem, that is, the moduli stabilization problem.

In both bottom-up and top-down approaches, the moduli stabilization  is important and necessary to determine the flavor structure. 
Unless these moduli fields are stabilized at a high scale larger enough than the current observational 
bound, the predictability of the modular flavor models is lost. 
For example, in Refs.~\cite{Kobayashi:2019xvz,Kobayashi:2019uyt} the possibilities of favorable moduli 
stabilization were studied by assuming modular symmetric non-perturbative effects.

One of powerful approaches to realize the moduli stabilization is flux compactifications, 
taking into account background field values of higher-dimensional tensor fields. 
For instance, in Type IIB flux compactifications, discrete modular symmetry is classified 
on toroidal background \cite{Kobayashi:2020hoc}, and spontaneous CP-violation is discussed in toroidal \cite{Kobayashi:2020uaj} and Calabi-Yau backgrounds~\cite{Ishiguro:2020nuf}. 
In this paper, we study how to fix the modulus value $\tau$ in the light of the moduli stabilization due to 
flux compactifications.

We deal with possible configurations of three-form fluxes in Type IIB string theory.
Then, we analyze systematically stabilized values of moduli, investigating their probabilities in the $SL(2,\mathbb{Z})$ 
fundamental region and showing which values are favorable from the moduli stabilization.
That is the so-called string landscape.
Then, we examine implications of our moduli stabilization from the viewpoint of 
modular flavor models. 
For illustrative purposes, we consider classes of modular $A_4$ models extensively studied in Ref.~\cite{Ding:2019zxk}. 
Then, we compare the predictions of these phenomenological models with our results in the string landscape. 
So far, the structure of Type IIB string landscape has been statistically studied in Refs. \cite{Ashok:2003gk,Douglas:2003um,Denef:2004ze} and 
in particular, the enhanced symmetries appear in particular points in moduli spaces of the complex 
structure moduli and the axio-dilaton \cite{DeWolfe:2004ns}. 
Such distributions of the moduli fields are useful to understand not only 
the nature of string landscape itself, 
but also phenomenological aspects of the 4D effective action. 
Indeed, the complex structure moduli determine the flavor structure of quarks and leptons on 
magnetized D-brane setup. 
It is expected that the probability distributions of complex structure moduli shed new light on the 
phenomenological approach. 
Our findings about the distribution of the complex structure modulus $\tau$ and the axio-dilaton $S$ are summarized as follows:

\begin{itemize}
\item
Distribution of the complex structure modulus $\tau$ determining the flavor structure of quarks and leptons is 
mostly clustered at smaller vacuum expectation values rather than the scattered distribution. 
Remarkably, the $\mathbb{Z}_3$ fixed point on the fundamental domain of the $SL(2,\mathbb{Z})$ moduli space is statistically favored by the weak string coupling and 
the tadpole cancellation condition. 
Also, distributions are clustered at other specific points such as intersecting points between $|\tau|^2=k/2$ and 
${\rm Re}\,\tau=0,\pm 1/4, \pm1/2$, although their probabilities are less than the $\mathbb{Z}_3$ fixed point.

\item
Distribution of the complex structure modulus $\tau$ depends on values of the axio-dilaton $S$, although 
the $\mathbb{Z}_3$ fixed point is universally favored in the landscape. 
For instance, the $\mathbb{Z}_2$ fixed point in the moduli space of the complex structure modulus $\tau$
is favored on the same $\mathbb{Z}_2$ fixed point in that of the axio-dilaton $S$.

\item
The CP-violating phase is determined by the vacuum expectation value of the complex structure modulus $\tau$.
We find that an axionic field of the complex structure modulus ${\rm Re}\,\tau$ is mostly stabilized 
at the CP-conserving vacua ${\rm Re}\,\tau=0,\pm 1/2$ or the CP-breaking vacua ${\rm Re}\,\tau=\pm 1/4$, 
although the CP-breaking vacua is statistically disfavored in the landscape. 
However, the CP-breaking vacua ${\rm Re}\,\tau=\pm 1/4$ are statically favored only if 
the axion ${\rm Re}\,S$ has certain values.
Thus, there is a strong correlation among CP phases in Yukawa couplings and the CP phase due to the axion ${\rm Re}\,S$.

\item
When we apply the distribution of the complex structure modulus to the modular $A_4$ models in Ref.~\cite{Ding:2019zxk}, 
one can predict the theoretically favorable vacuum expectation values 
of the moduli and its probability in the string landscape. 

\end{itemize}

This paper is organized as follows. 
First, we review the Type IIB supergravity action on $T^6/(\mathbb{Z}_2\times \mathbb{Z}^\prime_2)$ orientifold. 
The modular symmetries associated with 
the complex structure of tori and the axio-dilaton, and the stabilization of the moduli fields are 
reviewed in Section \ref{sec:2}. 
We next analyze the distributions of complex structure moduli in Section \ref{sec:3}, 
in which the phenomenologically applicable results are summarized. 
In Section \ref{sec:4}, we compare the phenomenological predictions of modular $A_4$ models 
with the distributions of the complex structure moduli. 
Finally, we conclude this paper in Section \ref{sec:con}. 
The detail of the numerical search is summarized in Appendix \ref{app}.

\section{Theoretical setup}
\label{sec:2}

In this section, we briefly review the phenomenologically attractive modular symmetry appearing in the string-derived  
effective action, with an emphasis on Type IIB supergravity on the factorizable $T^6$ torus 
and its orientifold. 
The dynamics of the moduli fields is discussed in the context of flux compactifications as 
reviewed in Section~\ref{subsec:2_2} on a simple $T^6/(\mathbb{Z}_2\times \mathbb{Z}^\prime_2)$ orientifold background. 
The reader who are interested in the distributions of the moduli fields may skip to Section \ref{sec:3}.

\subsection{Modular symmetry}
\label{subsec:2_1}

Let us consider the factorizable 6D torus $T^6=(T^2)_1\times (T^2)_2\times (T^2)_3$, each of which is defined by a complex plane divided by the lattice $\Lambda_i$. 
The geometrical symmetry of each torus is given by the $SL(2,\mathbb{Z})_i$ modular symmetry. 
When we represent the basis of the lattice $\Lambda_i$ by $(e_{x^i}, e_{y^i})$, 
the basis of the lattice transforms as
\begin{align}
    \begin{pmatrix}
     e_{y^i}^\prime\\
     e_{x^i}^\prime
    \end{pmatrix}
     =
   \begin{pmatrix}
    p_i & q_i\\
    s_i & t_i
    \end{pmatrix}
    \begin{pmatrix}
     e_{y^i}\\
     e_{x^i}
    \end{pmatrix}
    ,
\end{align}
where
\begin{align}
    R_i =
    \begin{pmatrix}
    p_i & q_i\\
    s_i & t_i
    \end{pmatrix}
\end{align}
are the element of $SL(2,\mathbb{Z})_i$ satisfying $p_i t_i -q_is_i=1$. 
Such a passive transformation of the basis induces the modular transformation 
of the complex structure $\tau_i$ as well as the coordinate $z_i$ with reference to $e_{x^i}$ as\cite{Kikuchi:2020nxn}
\begin{align}
    \tau_i=\frac{e_{y^i}}{e_{x^i}} &\rightarrow \tau_i^\prime=\frac{e_{y^i}^\prime}{e_{x^i}^\prime} =\frac{p_i\tau_i+q_i}{s_i\tau_i +t_i}=R_i(\tau_i),
    \nonumber\\
    z_i=\frac{u_i}{e_{x^i}} &\rightarrow z_i^\prime =\frac{u_i}{e_{x^i}^\prime}=\frac{z_i}{s_i\tau_i +t_i},
\end{align}
where $u_i$ denote the coordinates of the complex plane. 
The generators of each $SL(2,\mathbb{Z})_i$ are given by 
\begin{align}
    S = 
    \begin{pmatrix}
    0 & -1\\
    1 & 0
    \end{pmatrix}
    ,\quad
    T = 
    \begin{pmatrix}
    1 & 1\\
    0 & 1
    \end{pmatrix}
    ,
\end{align}
satisfying the following algebraic relations
\begin{align}
    S^2 = -\mathbb{I},\quad
    S^4=(ST)^3=\mathbb{I}.
\end{align}
These $S$- and $T$-transformations bring $\tau_i$ to 
the fundamental domain of the $SL(2,\mathbb{Z})_i$ moduli space:
\begin{align}
\left\{
-\frac{1}{2}\leq \tau_i \leq 0,\, |\tau_i|\geq 1\right\}\, 
\cup\, 
\left\{
0< \tau_i < \frac{1}{2},\,|\tau_i|>1 
\right\}
.
\label{eq:fund}
\end{align}

To see the modular symmetry in the effective action of complex structure moduli, 
let us introduce the three-form basis in $H^3(T^6,\mathbb{Z})$,\footnote{We follow the convention of Ref.~\cite{Kachru:2002he}.} 
\begin{align}
    \alpha_0 &=dx^1\wedge dx^2 \wedge dx^3,\quad \beta^0= dy^1\wedge dy^2 \wedge dy^3,
    \\ \notag
    \alpha_{1}&=dy^1\wedge dx^2 \wedge dx^3,\quad
    \beta^{1}=-dx^1\wedge dy^2\wedge dy^3,
    \\ \notag
    \alpha_{2}&=dy^2\wedge dx^3 \wedge dx^1,\quad
    \beta^{2}=-dx^2\wedge dy^3\wedge dy^1,
    \\ \notag
    \alpha_{3}&=dy^3\wedge dx^1 \wedge dx^2,\quad
    \beta^{3}=-dx^3\wedge dy^1\wedge dy^2,
    \label{eq:3formbasis}
    \end{align}
which satisfy the orientation 
\begin{align}
    \int_{T^6} \alpha_I \wedge \beta^J = \delta^{J}_I,
\end{align}
with $I,J=0,1,2,3$. 
Then, the 4D kinetic terms of the complex structure moduli are 
obtained by the dimensional reduction of 10D Einstein-Hilbert action. 
It results in the following K\"ahler potential:
\begin{align}
    K = -\ln\left(-i\int_{T^6} \Omega \wedge \bar{\Omega}\right) =-\ln\left(i (\tau_1-\bar{\tau}_1)(\tau_2-\bar{\tau}_2)(\tau_3-\bar{\tau}_3) \right) ,    
    \label{eq:Kcs}
\end{align}
where we define the holomorphic three-form on $T^6$ as
\begin{align}
    \Omega =dz_1\wedge dz_2 \wedge dz_3.
\end{align}

The modular symmetries of the complex structure moduli are checked to see the K\"ahler-invariant quantity,
\begin{align}
G= K + \ln |W|^2,
\end{align}
which consists of the K\"ahler potential and the superpotential characterizing the 4D effective potential. 
Since the K\"ahler potential of the complex structure moduli transforms under the 
modular symmetry as
\begin{align}
K \rightarrow K + \sum_{i=1}^3 \ln |s_i \tau_i +t_i|^2, 
\end{align}
the effective potential is modular invariant up to the K\"ahler transformation, only 
if the superpotential has the modular weight 1 under each $SL(2,\mathbb{Z})_i$ modular symmetry, 
that is,
\begin{align}
W \rightarrow \frac{W}{\Pi_{i=1}^3 (s_i\tau_i +t_i)}.
\end{align}

So far, there exists the 4D ${\cal N}=8$ supersymmetry originating from the 10D ${\cal N}=2$ supersymmetry 
in the case of Type II string theory. 
To reduce ${\cal N}=8$ supersymmetry to ${\cal N}=1$ supersymmetry, we impose the orientifold and orbifold projections on $T^6$ background. 
Especially, we focus on ${\cal M}=T^6/(\mathbb{Z}_2\times \mathbb{Z}_2^\prime)$ orientifold throughout this paper. 
The $\mathbb{Z}_2\times \mathbb{Z}_2^\prime$ orbifold projections are chosen as
\begin{align}
\theta : (z_1, z_2, z_3)\rightarrow (-z_1, -z_2, z_3),\quad
\theta^\prime : (z_1, z_2, z_3)\rightarrow (z_1, -z_2, -z_3),
\end{align}
and the orientifold projection is specified by
\begin{align}
{\cal R}: (z_1, z_2, z_3)\rightarrow (-z_1, -z_2, -z_3).
\end{align}
We denote 
the world-sheet parity projection by $\Omega_p$ and the left-moving fermion number operator by $F_L$, respectively.
It is known that under the latter two operators $\Omega_p (-1)^{F_L}$, the metric and the axio-dilaton are even 
and other higher-dimensional form fields such Kalb-Ramond field $B$ and Ramond-Ramond (RR) fields $C_{2p}$ 
behave as 
\begin{align}
\Omega_p (-1)^{F_L} B = -B,\quad
\Omega_p (-1)^{F_L} C_{2p} = (-1)^p C_{2p}.
\end{align}
Note that all three-form basis in Eq.~(\ref{eq:3formbasis}) are invariant under the orbifold projections, 
and furthermore, these bases are odd under the ${\cal R}$. 
Hence, we still have the three complex structure moduli, that is the  $h^{2,1}_-({\cal M}) ={\rm dim}H^{2,1}_-({\cal M})=3$, 
where the cohomology groups of tori split into $H^{2,1}({\cal M})=H^{2,1}_+({\cal M}) \oplus H^{2,1}_-({\cal M})$ 
under the orientifold projection. 
The kinetic terms of the complex structure moduli are also described in Eq.~(\ref{eq:Kcs}) 
in the same way as the factorizable $T^6$ background.

\subsection{Flux compactifications on $T^6/(\mathbb{Z}_2\times \mathbb{Z}_2^\prime)$ orientifold}
\label{subsec:2_2}

The moduli stabilization is inevitable to discuss the flavor structure of quarks and leptons.
In particular, the flux compactification is useful to determine the vacuum expectation 
values of the moduli fields in a controlled way. 
In this paper, we focus on $T^6/(\mathbb{Z}_2\times \mathbb{Z}_2^\prime)$ orientifold 
as explained in the previous section. 

On the $T^6/(\mathbb{Z}_2\times \mathbb{Z}_2^\prime)$ background, 
the effective K\"ahler potential of the moduli fields is described by
\begin{align}
    K = -\ln (-i(S -\Bar{S})) -2\ln {\cal V} -\ln\left(i (\tau_1-\bar{\tau}_1)(\tau_2-\bar{\tau}_2)(\tau_3-\bar{\tau}_3) \right),     
    \label{eq:Keff}
\end{align}
where we include the axio-dilaton $S$ and the volume of the torus ${\cal V}$ 
in the Einstein-frame measured in units of the string length $l_s=2\pi \sqrt{\alpha^\prime}$. 
Here and in what follows, we adopt the reduced Planck mass unit $M_{\rm Pl}=1$. 
The effective action has three-types of $SL(2,\mathbb{Z})$ symmetry, namely 
$SL(2,\mathbb{Z})_S$ for the axio-dilaton, $SL(2,\mathbb{Z})_T$ for the volume 
moduli and $SL(2,\mathbb{Z})_\tau$ for the complex structure moduli.\footnote{The 
fundamental region of $SL(2,\mathbb{Z})_{S,T}$ is the same with Eq. (\ref{eq:fund}) by replacing 
$\tau_i$ with those moduli fields.}
In Type IIB string setup with D3/D7-branes, the complex structure moduli is of particular interest 
for the flavor structure of quarks and leptons localized on D-branes. 
These complex structure moduli and the axio-dilaton can be stabilized in the context of flux compactifications.

In the context of Type IIB string theory, there exists the kinetic term of three-form $G_3$ 
consisting of Ramond-Ramond (RR) $F_3$ and Neveu-Schwarz (NS) three-forms $H_3$, namely 
$G_3=F_3-SH_3$. After the dimensional reduction of this kinetic term on $T^6/(\mathbb{Z}_2\times \mathbb{Z}_2^\prime)$ background and taking into account background fluxes of these three-forms, 
one can obtain the moduli-dependent superpotential in the four-dimensional effective action~\cite{Gukov:1999ya}
\begin{align}
    W = \frac{1}{l_s^2}\int G_3 \wedge \Omega.
    \label{eq:Weff}
\end{align}
The RR and NS three-forms are expanded on the basis of Eq.~(\ref{eq:3formbasis});
\begin{align}
\frac{1}{l_s^2}F_3 &= a^0 \alpha_0 +a^{i}\alpha_{i} +b_{i}\beta^{i} +b_0 \beta^0,
\nonumber\\
\frac{1}{l_s^2}H_3 &= c^0 \alpha_0 +c^{i}\alpha_{i} +d_{i}\beta^{i} +d_0 \beta^0,
\label{eq:F3H3_T6Z2}
\end{align}
where we denote the integral flux quanta by $\{a^{0,1,2,3},b_{0,1,2,3},c^{0,1,2,3},d_{0,1,2,3}\}$. 
On $T^6/(\mathbb{Z}_2\times \mathbb{Z}_2^\prime)$ geometry, these flux quanta are restricted 
to be in multiples of 4 or 8 depending on the existence of discrete torsion \cite{Vafa:1986wx,Vafa:1994rv,Douglas:1998xa}. 
Throughout this paper, we analyze the latter case, since one can focus on the dynamics of 3 
untwisted complex structure moduli, 
denoted by $\tau_{1,2,3}$ in the effective action. 
Then, the explicit form of the superpotential is given by
\begin{align}
    W &= a^0 \tau_1\tau_2\tau_3  +c^1 S \tau_2\tau_3
    +c^2 S\tau_1\tau_3 +c^3 S\tau_1\tau_2
 - \sum_{i=1}^3 b_i \tau_i + d_0 S
 \nonumber\\
& -c^0 S\tau_1\tau_2\tau_3 - a^1 \tau_2\tau_3
    - a^2\tau_1\tau_3 - a^3\tau_1\tau_2
+ \sum_{i=1}^3 d_i S \tau_i - b_0,
\label{eq:W}
\end{align}
which generates the potential of the axio-dilaton and complex structure moduli. 
Indeed, by combining the superpotential (\ref{eq:W}) and the K\"ahler potential (\ref{eq:Keff}), 
the 4D scalar potential is constructed in the framework of 4D ${\cal N}=1$ supergravity. 
Since the superpotential is independent of the K\"ahler moduli, 
the 4D scalar potential can be simplified as
\begin{align}
    V = e^K \left(K^{I\bar{J}}D_IW D_{\bar{J}}\bar{W} \right),
\end{align}
with $I,J=S,\tau_1,\tau_2,\tau_3$. 
Here, we denote the inverse of the K\"ahler metric $K_{I\bar{J}}$ by $K^{I\bar{J}}$ as computed from the 
K\"ahler potential $K_{I\bar{J}}=\partial_I\partial_{\bar{J}}K$, and the covariant derivative of the superpotential is given by
\begin{align}
    D_I W = \partial_I W +(\partial_I K)W.
\end{align}
Note that the negative term $-3|W|^2$ in the scalar potential is cancelled by 
the no-scale structure of the K\"ahler moduli.

Before going into the detail to analyze the structure of the scalar potential, 
we discuss the symmetry possessed in the flux-induced effective action. 
To simplify our analysis, we concentrate on the isotropic torus, 
where the complex structure moduli are constrained on the following locus:
\begin{align}
\tau \equiv \tau_1=\tau_2 =\tau_3,
\end{align}
and correspondingly, flux quanta are redefined as
\begin{align}
a\equiv a^1=a^2=a^3,\quad b\equiv b_1=b_2=b_3,\quad
c\equiv c^1=c^2=c^3,\quad d\equiv d_1=d_2=d_3,
\end{align}
meaning that we have totally 8 independent flux quanta; $\{a^0,a,b,b_0 \}$ for RR flux 
and $\{c^0,c,d,d_0 \}$ for NS flux.  
This system has three types of symmetries as shown below. 

First, as mentioned in Section~\ref{subsec:2_1}, the effective action is invariant under 
the  $SL(2,\mathbb{Z})_\tau\equiv SL(2,\mathbb{Z})_1= SL(2,\mathbb{Z})_2= SL(2,\mathbb{Z})_3$ modular symmetry only if the superpotential
\begin{align}
    W &= a^0 \tau^3- 3a\tau^2- 3b \tau-b_0 -S\left(c^0\tau^3-3c\tau^2-3d\tau-d_0 \right),
\label{eq:Wsim}
\end{align}
has the modular weight 3, which is different from 1 owing to the identification of three complex structure moduli. 
Under $S$- and $T$-transformations of $SL(2,\mathbb{Z})_\tau$ modular group, 
the effective action is indeed modular invariant when the flux quanta transform as~\cite{Betzler:2019kon},

\begin{itemize}
\item $S: \tau \rightarrow -1/\tau$
\begin{align}
a^0\rightarrow b_0,\quad a\rightarrow b,\quad b\rightarrow -a,\quad b_0 \rightarrow -a^0,
\\ \notag
c^0\rightarrow d_0,\quad c\rightarrow d,\quad d\rightarrow -c,\quad d_0 \rightarrow -c^0.
\end{align}
\item $T: \tau \rightarrow \tau+1$
\begin{align}
&a^0\rightarrow a^0,\quad a\rightarrow a + a^0,\quad b\rightarrow b-2a-a^0,\quad b_0 \rightarrow b_0-3b +3a +a^0,
\\ \notag
&c^0\rightarrow c^0,\quad c\rightarrow c + c^0,\quad d\rightarrow d-2c-c^0,\quad d_0 \rightarrow d_0-3d+3c+c^0.
\end{align}
\end{itemize}

Second, the axio-dilaton enjoys the $SL(2,\mathbb{Z})_S$ modular symmetry. 
Under the $SL(2,\mathbb{Z})_S$ transformation of the axio-dilaton
\begin{align}
    S \rightarrow S^\prime =\frac{pS+q}{sS +t}=R_S(\tau_i),
\end{align}
with
\begin{align}
    R_S =
    \begin{pmatrix}
    p & q\\
    s & t
    \end{pmatrix},
\end{align}
being the element of $SL(2,\mathbb{Z})_S$ satisfying $p t -qs=1$, 
the effective action is modular invariant under the following transformation of the RR and NS fluxes,
\begin{align}
    \begin{pmatrix}
    F\\
    H
     \end{pmatrix}
     \rightarrow
    \begin{pmatrix}
    p & q\\
    s & t
    \end{pmatrix}
     \begin{pmatrix}
    F\\
    H
     \end{pmatrix}   
.
\end{align}
Specifically, the flux quanta transform under the $S$- and $T$-transformations of $SL(2, \mathbb{Z})_S$ as
\begin{itemize}
    \item $S: S \rightarrow -1/S$ 
    \begin{align}
&a^0\rightarrow -c^0,\quad a\rightarrow -c,\quad b\rightarrow -d,\quad b_0 \rightarrow -d_0,
\\ \notag
&c^0\rightarrow a^0,\quad c\rightarrow a,\quad d\rightarrow b,\quad d_0 \rightarrow b_0.
\end{align}
\item $T: S \rightarrow S+1$
\begin{align}
&a^0\rightarrow a^0+c^0,\quad a\rightarrow a+c,\quad b\rightarrow b+d,\quad b_0 \rightarrow b_0+d_0,
\\ \notag
&c^0\rightarrow c^0,\quad c\rightarrow c,\quad d\rightarrow d,\quad d_0 \rightarrow d_0.
    \end{align}
\end{itemize}

Third, the effective action is invariant under flipping the overall sign of flux quanta, 
namely
\begin{align}
(a^0, a, b, b_0, c^0, c, d, d_0)
\rightarrow
-(a^0, a, b, b_0, c^0, c, d, d_0),
\label{eq:third}
\end{align}
which changes the sign of the superpotential, $W\rightarrow -W$, but the effective action is invariant, taking into account 
the K\"ahler transformation. This symmetry can be used to reduce $SL(2, \mathbb{Z})_{\tau, S}$ to $PSL(2, \mathbb{Z})_{\tau, S}$.
Then, these three symmetries are employed to count the physically-distinct vacua as 
analyzed in Section \ref{sec:3}.

Finally, we comment on constraints for the flux quanta. 
These flux quanta induce the D3-brane charge through the Bianchi identity of the RR field,
\begin{align}
    N_{\rm flux}&= \frac{1}{l_s^4}\int H_3\wedge F_3 = c^0b_0 -d_0a^0 +\sum_i (c^ib_i -d_ia^i) = 
    c^0b_0 -d_0a^0 +3(cb -da),
    \label{eq:nD3}
\end{align}
which should be cancelled by the contributions of D3/D7-branes and O3/O7-planes\footnote{Note that $N_{\rm flux}$ remains invariant under $SL(2, \mathbb{Z})_{\tau, S}$ and the sign flip (\ref{eq:third}).}. 
It was known that $N_{\rm flux}$ are bounded as $0\leq N_{\rm flux}\leq {\cal O}(10)$ 
for some explicit setups on the toroidal $T^6/\mathbb{Z}_M$ and $T^6/(\mathbb{Z}_M\times \mathbb{Z}_N)$ 
orientifolds~\cite{Lust:2006zh}. 
The positivity of $N_{\rm flux}$ is a consequence of the supersymmetry. 
When we uplift the Type IIB orientifolds to its strong coupling regime, namely 
F-theory, O3-plane contributions are encoded in the Euler number of Calabi-Yau fourfolds, 
whose largest value is known as $1820448$~\cite{Candelas:1997eh,Taylor:2015xtz}. In this case, the flux-induced D3-brane charge 
is bounded as
\begin{align}
0\leq N_{\rm flux}\leq N_{\rm flux}^{\rm max}={\cal O}(10^5).
    \label{eq:Nflux}
\end{align}
Similar to the analysis in Ref.~\cite{Betzler:2019kon}, we adopt this bound (\ref{eq:Nflux}) in the following 
analysis, where we examine the distributions of moduli fields by changing the maximum value of 
the flux-induced D3-brane charge, $N_{\rm flux}^{\rm max}$. 
Since the flux quanta are in multiple of 8 on $T^6/(\mathbb{Z}_2\times \mathbb{Z}_2^\prime)$ geometry, 
$N_{\rm flux}$ should be in multiple of 192, namely\footnote{See for more details, Appendix \ref{app}.}
\begin{align}
N_{\rm flux}\in 192\,\mathbb{Z}.
\end{align}
\subsection{Supersymmetric minima}
\label{subsec:2_3}

In this section, we consider the supersymmetric minimum solutions of the moduli fields 
on the isotropic torus.

The supersymmetric minimum conditions of the moduli fields are 
given by
\begin{align}
\partial_{\tau} W =0,\quad \partial_{S}W=0,\quad W=0,
\label{eq:SUSY-min}
\end{align}
where the third  condition is due to the supersymmetric condition for K\"ahler moduli\footnote{Since we consider the no-scale scalar potential in this paper, we cannot stabilize K\"ahler moduli here. Actually, this is the condition at the classical level so we should adopt a stabilizing mechanism of them. }.  

First of all, we separate the superpotential into RR and NS parts as $W=W_{\rm RR}-SW_{\rm NS}$, where
\begin{align}
    W_{\rm RR} = a^0 \tau^3- 3a\tau^2- 3b \tau-b_0, \\
    W_{\rm NS} = c^0\tau^3-3c\tau^2-3d\tau-d_0.
\end{align}

The minimum conditions $\partial_{S} W = 0$ and $W=0$ lead
\begin{align}
    W_{\rm RR} = 0 = W_{\rm NS}.
\end{align}
Each of the equations, $W_{\rm RR}=0$ and $W_{\rm NS}=0$,  is the cubic equation with respect to $\tau$ and the coefficients are real. 
The stabilized value ${\rm Im}\,\tau$ must be positive, i.e. ${\rm Im}\,\tau > 0$; otherwise, the volume of $T^6$ is vanishing.
To find a solution $\tau$ with its non-zero imaginary part, the cubic equation must have two common complex solutions which are conjugate with each other and a real one which need not be common. 
Taking into account them, $W_{\rm RR}$ and $W_{\rm NS}$ should be factorizable as\footnote{In the following section, we follow the notation used in Ref. \cite{Betzler:2019kon} and $\{m, l, n, u, v, s, r\}$ are real numbers which are given by the flux quanta $\{a^0, a, b, b_0\}$ or $\{c^0, c, d, d_0\}$. Moreover, we can make $\{m, l, n, u, v, s, r\}$ be integers by using Gauss's lemma. More details are given in Appendix \ref{app}. }
\begin{align}
    W_{\rm RR} = (r \tau + s) P(\tau), \\
    W_{\rm NS} = (u \tau + v) P(\tau),
\end{align}
for a quadratic (integer-coefficient) polynomial $P(\tau)$. 
It indicates that the complex $\tau$ is obtained from $P(\tau)=0$. 

Next, let us consider the condition $\partial_{\tau} W =0 $ whose explicit form is given by
\begin{align}
    P(\tau)\partial_\tau\{(r \tau + s) - S (u \tau + v) \} + \{(r \tau + s) - S (u \tau + v) \} \partial_\tau P(\tau) = 0.
\end{align}
For a complex solution $\tau$ given by $P(\tau)=0$, it is required to satisfy $\partial_\tau P(\tau) = 0$ or 
\begin{align}
    S =  \frac{ r \tau + s}{u \tau + v}.
    \label{eq:Svev}
\end{align}
Indeed, $\partial_\tau P(\tau) = 0$ leads a real $\tau$ so that the vacuum expectation value of $S$ is given as above.\footnote{Here and in what follow, we omit the symbol of the vacuum expectation values.} 

To obtain an explicit form of $\tau$, we set 
\begin{align}
    P(\tau) = l \tau^2 + m \tau + n ,
\end{align}
for $m^2 - 4 l n < 0$, and then we arrive at the following vacuum expectation value of $\tau$:
\begin{align}
    \tau &= \frac{ - m + \sqrt{m^2 - 4 l n}}{2 l} \quad (l, n > 0),
    \label{eq:lnpositive}
    \\
    \tau &= \frac{ - m - \sqrt{m^2 - 4 l n}}{2 l} \quad (l, n < 0).
    \label{eq:lnnegative}
\end{align}


As we reviewed in Section \ref{subsec:2_2}, there are two symmetries for 
$S$, $\tau$, namely $SL(2, \mathbb{Z})_S$, $SL(2, \mathbb{Z})_\tau$ which leave the effective action invariant.  
Hence, there exists an equivalence relation between two given flux vacua, 
which can be found by checking the invariance of 
supersymmetric conditions under the modular transformations. 
For the detail about the equivalence, see, Ref. \cite{Betzler:2019kon}. 
However, to be self-contained, we briefly summarize such an equivalence in Appendix \ref{app}. 
We have to be careful not to make double-counting there.

\section{Distributions of the moduli fields}
\label{sec:3}

In this section, we systematically analyze the distributions of complex structure modulus $\tau$ 
as well as the axio-dilaton $S$ by the moduli stabilization through flux compactifications as 
explained in the previous section.
We set  the isotropic torus $\tau=\tau_1=\tau_2=\tau_3$. 
On the basis of the setup in the previous section, 
we show  our results on  distributions of the complex structure modulus $\tau$ and the axio-dilaton $S$. 
In Section \ref{subsec:3_1}, we first investigate the distributions of stable vacua on the plane of complex structure modulus $\tau$,  taking into account the upper bound for the flux-induced D3-brane charge $N_{\rm flux}^{\rm max}$ in Eq.~(\ref{eq:Nflux}), 
which fixes the maximal size of fluxes, and the lower bound for the string coupling $g_s={\rm Im}\,S^{-1}$. 
These results indicate the correlation between the distributions of the complex structure modulus 
and the axio-dilaton. 
We next deal with the distributions of stable vacua on the plane of the axio-dilaton in Section \ref{subsec:3_2}, 
especially we pick up statistically favored vacuum expectation values of the axio-dilaton and discuss its consequence for the distribution of the complex structure modulus. 
Section \ref{subsec:3_3} is devoted to analyze the CP-breaking vacua.

\subsection{Distribution of complex structure modulus without fixing the axio-dilaton}
\label{subsec:3_1}

We analyze the structure of the supersymmetric vacua as analytically demonstrated in Section \ref{subsec:2_3}. 
We search all the possible flux configurations $(a^0, a, b, b_0, c^0, c, d, d_0)$, which satisfy the tadpole cancellation 
condition (\ref{eq:Nflux}) for each\footnote{Here and in what follows, we omit the factor 192 unless we specify it.}
\begin{align}
N_{\rm flux}^{\rm max}=10,100,250,500,1000. 
\end{align}
By use of fluxes, we examine stabilized moduli values, satisfying the minimum conditions 
(\ref{eq:SUSY-min}), i.e., Eqs. (\ref{eq:Svev}) and $P(\tau)=0$. 
Since the effective action has three types of symmetries as mentioned in Section \ref{subsec:2_2}, 
it is required not to  double count the number of stable vacua. 
As a result, we find the number of physically-distinct stable vacua as summarized in Table \ref{tab:Number}. The number of stable vacua increases as we allow the larger upper bound for the flux-induced D3-brane 
charge $N_{\rm flux}^{\rm max}$. 
\begin{table}[H]
\centering
\begin{tabular}{|c|c|c|c|c|c|} \hline
 & $N_{\rm flux}^{\rm max}=10$ & $N_{\rm flux}^{\rm max}=100$ & $N_{\rm flux}^{\rm max}=250$ 
 & $N_{\rm flux}^{\rm max}=500$ & $N_{\rm flux}^{\rm max}=1000$ \\ \hline
$\#$ of stable vacua & 312 & 29218 & 178191 & 720710 & 2896221 \\ \hline
\end{tabular}
\caption{Number of stable vacua for each $N_{\rm flux}^{\rm max}$ determining the range of flux quanta 
through the tadpole cancellation condition (\ref{eq:Nflux}).} 
\label{tab:Number}
\end{table}
The number of our finding stable vacua behaves as\footnote{Here we distinguish two vacua whose corresponding flux configuration is different from each other, up to the $SL(2, \mathbb{Z})$ transformation. As discussed in Appendix \ref{app}, we first fix degree of freedom of $T$-transformation and next divide it by $S$-transformation. 
Note that these procedures to count the number of stable vacua is different from Ref. \cite{Betzler:2019kon}.}
\begin{align}
    \text{\# of stable vacua} \simeq 2.89527 \left( \frac{ N_{\rm flux}^{\rm max} }{ 192 }\right)^2.
\end{align}

\begin{figure}[H]
\begin{minipage}{0.5\hsize}
  \begin{center}
  \includegraphics[height=120mm]{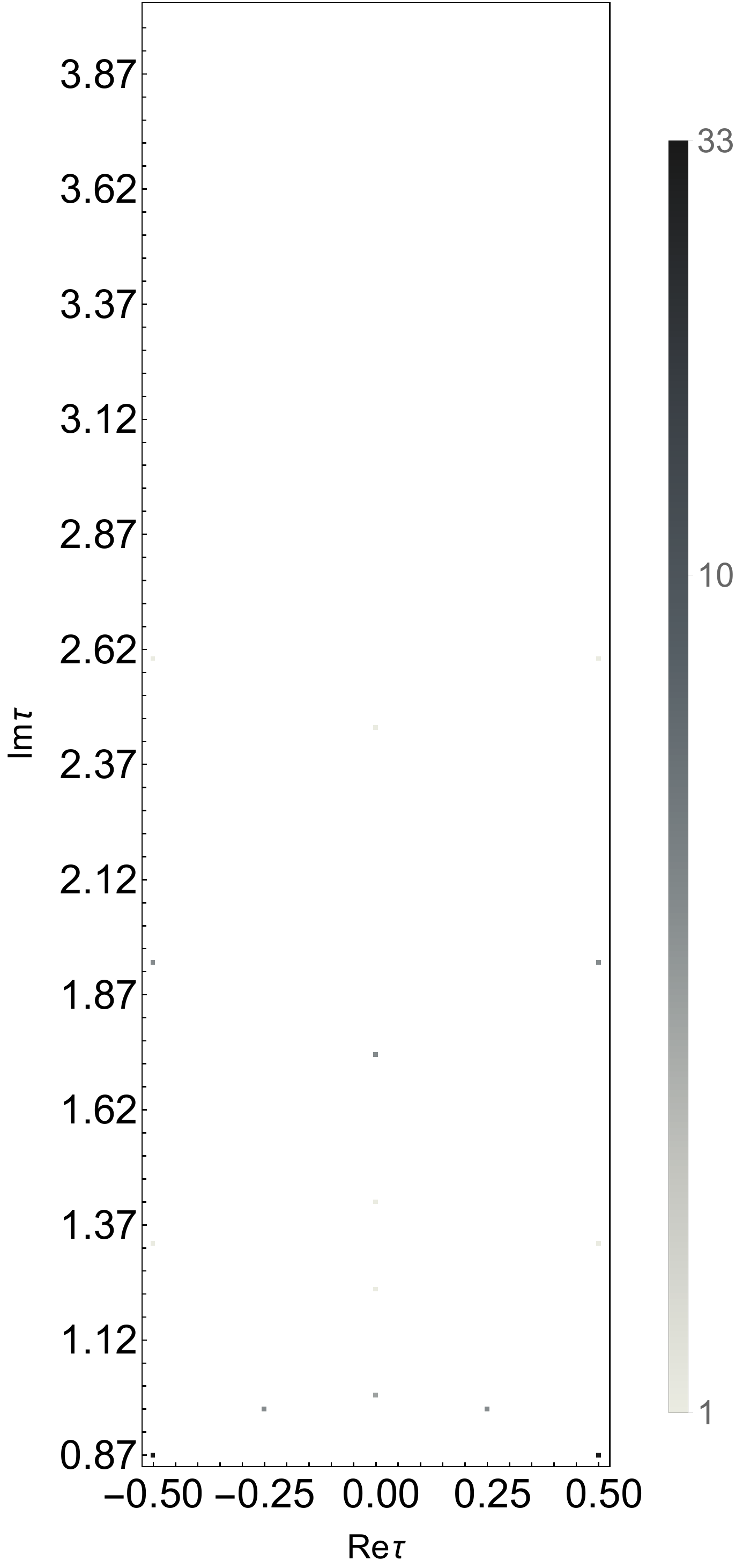}
  \end{center}
 \end{minipage}
 \begin{minipage}{0.5\hsize}
  \begin{center}
   \includegraphics[height=120mm]{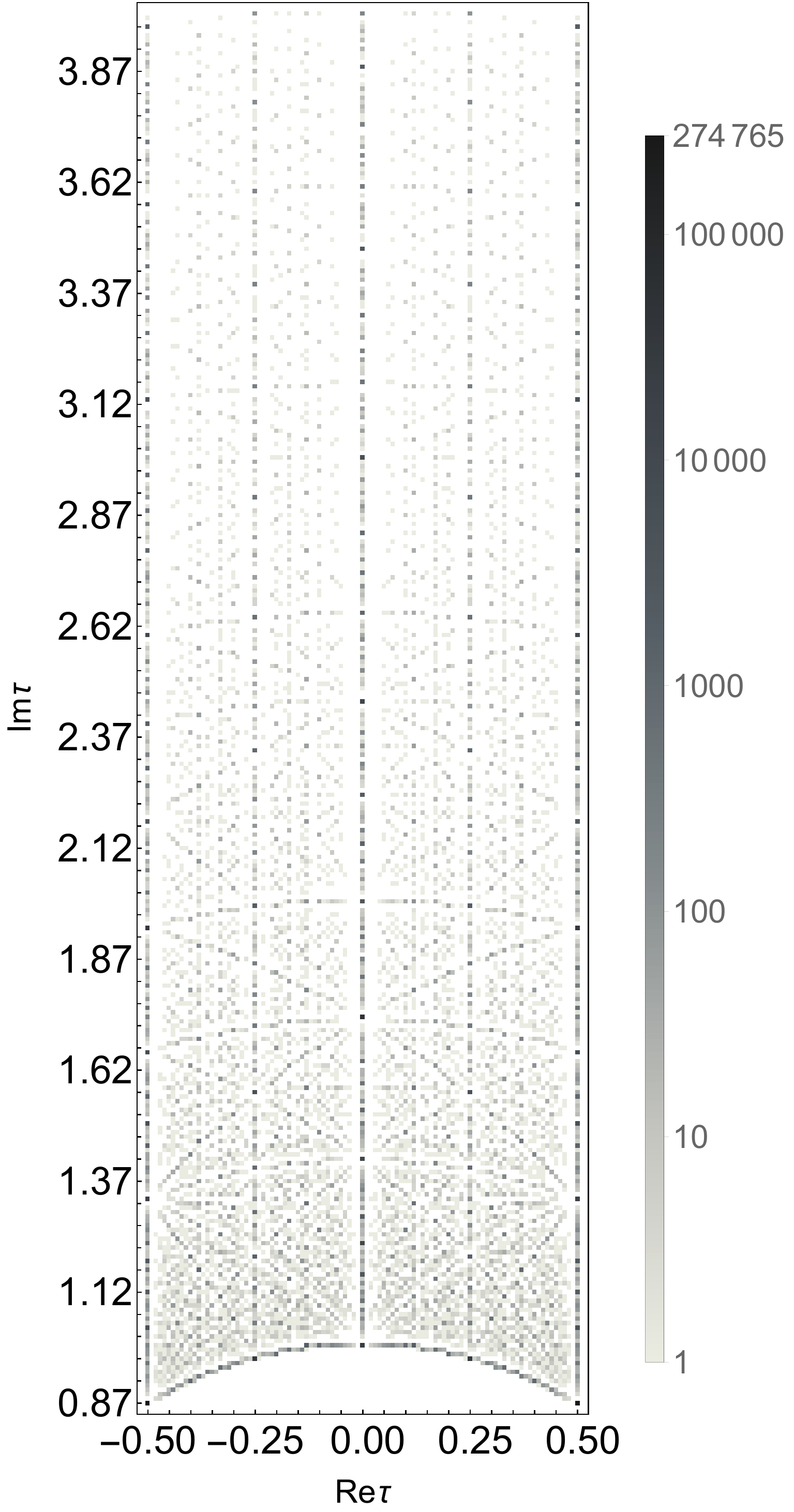}
  \end{center}
 \end{minipage}
  \caption{The numbers of stable vacua on $({\rm Re}\,\tau, {\rm Im}\,\tau)$ plane for $N_{\rm flux}^{\rm max}=10$ 
  in the left panel and $N_{\rm flux}^{\rm max}=1000$ in the right panel, respectively.}
\label{fig:1}
\end{figure}

In this section, we first focus on the distribution of the complex structure modulus $\tau$ without 
imposing any bound for the axio-dilaton $S$. 
The distribution of the complex structure modulus $\tau$ in the fundamental region of the $SL(2,\mathbb{Z})_\tau$ 
moduli space (\ref{eq:fund}) is drawn in Figure \ref{fig:1} for the case with 
$N_{\rm flux}^{\rm max}=10$ in the left panel and $N_{\rm flux}^{\rm max}=1000$ in the right panel, respectively. 
We find that the stable vacua are clustered in the dark color region in both panels 
and especially, most of the stable vacua are clustered at smaller vacuum expectation values of 
${\rm Im}\,\tau$ rather than the scattered distribution. 
That is a consequence of the condition for the flux quanta through the tadpole cancellation condition (\ref{eq:Nflux}). 

Furthermore, the $\mathbb{Z}_3$ fixed point of the $SL(2,\mathbb{Z})$ modular group is favored in the landscape. 
We recall that the fixed points $\tau_{\rm fix}$ in the fundamental region of the $SL(2,\mathbb{Z})$ moduli space 
are defined as
\begin{align}
\gamma_0 \tau_{\rm fix}=\tau_{\rm fix},
\end{align}
where $\gamma_0$ is the element of $SL(2,\mathbb{Z})$ transformation called as a stabilizer of $\tau_{\rm fix}$. 
It is known that there exist three fixed points in the fundamental region such as
\begin{align}
\tau_{\rm fix}=i,\quad i\infty, \quad w,
\end{align}
with $w=\frac{-1+i\sqrt{3}}{2}$, where the first two points and the latter one correspond to the $\mathbb{Z}_2$ and $\mathbb{Z}_3$ fixed points, 
respectively. 

\begin{table}[H]
\scalebox{0.76}{
\begin{tabular}{|c|c|c|c|c|c|c|c|c|c|c|} \hline
$({\rm Re}\,\tau, {\rm Im}\,\tau)$ & ($-\frac{1}{2}, \frac{\sqrt{3}}{2}$) & ($0, \sqrt{3}$) & ($-\frac{1}{2}, \frac{\sqrt{15}}{2}$) 
 & ($-\frac{1}{4}, \frac{\sqrt{15}}{4}$) &($0, 1$) & ($-\frac{1}{2}, \frac{3\sqrt{3}}{2}$) & ($0, \sqrt{\frac{3}{2}}$) & (0, $\sqrt{6}$) & ($0, \sqrt{2}$) & ($-\frac{1}{2}, \frac{\sqrt{7}}{2}$)  \\ \hline
Probability ($\%$) & 62.3 & 7.55 & 7.55 & 7.55 & 5.66 & 1.89 & 1.89 & 1.89 & 1.89 & 1.89 \\ \hline
\end{tabular}
}
\caption{Probabilities of the stable vacua as functions of $({\rm Re}\,\tau, {\rm Im}\,\tau)$ 
in the descending order of the probability for $N_{\rm flux}^{\rm max}=10$.} 
\label{tab:UNflux10}
\end{table}
\begin{table}[H]
\scalebox{0.76}{
\begin{tabular}{|c|c|c|c|c|c|c|c|c|c|c|} \hline
$({\rm Re}\,\tau, {\rm Im}\,\tau)$ & ($-\frac{1}{2}, \frac{\sqrt{3}}{2}$) & ($0, \sqrt{3}$) & ($-\frac{1}{2}, \frac{\sqrt{15}}{2}$) 
 & ($-\frac{1}{4}, \frac{\sqrt{15}}{4}$) &($0, 1$) & ($-\frac{1}{2}, \frac{\sqrt{7}}{2}$) & ($0, \sqrt{6}$) & (0, $\sqrt{\frac{3}{2}}$) & ($0, \sqrt{2}$) & ($-\frac{1}{2}, \frac{3\sqrt{3}}{2}$) \\ \hline
Probability ($\%$) & 40.3 & 7.55 & 4.85 & 4.85 & 3.79 & 2.43 & 1.88 & 1.88 & 1.88  & 1.49\\ \hline
\end{tabular}
}
\caption{Probabilities of the stable vacua as functions of $({\rm Re}\,\tau, {\rm Im}\,\tau)$ 
in the descending order of the probability for $N_{\rm flux}^{\rm max}=1000$.} 
\label{tab:UNflux1000}
\end{table}

Indeed, we calculate the probability of stable vacua as shown in Tables \ref{tab:UNflux10} 
and \ref{tab:UNflux1000}.\footnote{Here and in what follows, we add ${\rm Re} X = 1/2$ $(X = \tau, S)$ lines and the right side of unit circles to make the figures symmetric. However, to define probabilities correctly, we would not do the same thing on the tables including the probabilities. In addition, we divided 2D moduli space into domains whose size are $(0.01 \times 0.01)$ to draw the figures, but this configuration still captures fine structures such as voids.} 
These results indicate that the stable vacua are clustered at the $\mathbb{Z}_3$ fixed point, 
when $N_{\rm flux}^{\rm max}$ is small. 
Note that the smallness of $N_{\rm flux}^{\rm max}$ is more theoretically controllable, because we require the 
extension of Type IIB string setup to the F-theory extension in $N_{\rm flux}^{\rm max}\sim {\cal O}(10^3)$ regime. 
In addition to the $\mathbb{Z}_3$ fixed point, there exist other favorable modulus values such as 
the $\mathbb{Z}_2$ fixed point $({\rm Re}\,\tau, {\rm Im}\,\tau)=(0, 1)$ and $({\rm Re}\,\tau, {\rm Im}\,\tau)=(0, \sqrt{k})$ 
with $k=2,3,6$. 
All of these are on boundary of the $GL(2,\mathbb{Z})$ fundamental region.
Furthermore, all of them satisfy
\begin{align}
|\tau|^2=\frac{k}{2},
\end{align}
where $k$ is positive integer.
That is, all of the points in Tables \ref{tab:UNflux10} and \ref{tab:UNflux1000} except $({\rm Re}\,\tau, {\rm Im}\,\tau)=(-\frac{1}{4},\frac{\sqrt{15}}{4})$ are intersecting points between $|\tau|^2=k/2$ and ${\rm Re}\,\tau=0,-1/2$.\footnote{
In Ref.~\cite{Kobayashi:2020uaj}, it was shown that the rescaled modulus $\phi$ satisfies $|\phi|^2=1$ for CP-symmetric 
superpotential.}

In addition to the points in Tables \ref{tab:UNflux10} and \ref{tab:UNflux1000}, the following points:
\begin{align}
({\rm Re}\,\tau, {\rm Im}\,\tau)=\left(\pm \frac14,\frac{\sqrt{8k-1}}{4}\right)=\left(\pm \frac14,\frac{\sqrt{23}}{4}\right), \left(\pm \frac14,\frac{\sqrt{31}}{4}\right), 
\left(\pm \frac14,\frac{\sqrt{39}}{4}\right), \cdots, 
\label{eq:CPV-min}
\end{align}
 have ${\cal O}(0.1\mathchar`-1) \%$ of probability.
These points are also on the circles with the radii $\sqrt{k/2}$ including $({\rm Re}\,\tau, {\rm Im}\,\tau)=(\pm 1/4, \sqrt{15}/4)$.
These points are important from the viewpoint of CP violation, as will be studied in 
Section \ref{subsec:3_3}.

In general, from the analytic expression of $\tau$ in Eqs. (\ref{eq:lnpositive}) 
and (\ref{eq:lnnegative}), one can see that
\begin{align}
    |\tau|^2 =\frac{n}{l}, \label{eq:circle radius of tau}
\end{align}
i.e. all the vacuum expectation values of $\tau$ in the upper half-plane are on the semicircles and each radius of them is given by Eq. (\ref{eq:circle radius of tau}). 
Note that this does not mean that each semicircle becomes a solution to Eq. (\ref{eq:SUSY-min}) in its entirety.
Indeed, the solutions are discrete in the first place. 
If we assume that the vacua are distributed in those circles, it is needed to fold back into the region by using $T$-transformations since these circles protrude from the fundamental region.
This phenomenon can be seen from the pattern of Fig. \ref{fig:1}, and it implies that the overlap points including $(\frac{1}{4}, \frac{\sqrt{8k-1}}{4}), (\frac{1}{2}, \frac{\sqrt{4k-1}}{2})$ and $(0, \sqrt{k})$ listed above are statistically favored. 

Our results are in agreement with the statistical approach developed in Refs. \cite{Ashok:2003gk,Douglas:2003um,Denef:2004ze}, in which 
the discrete flux quanta are taken to be continuous one. As demonstrated on the toroidal background \cite{DeWolfe:2004ns}, the number of $W=0$ supersymmetric vacua with 
fixed complex structure modulus behaves as
\begin{align}
    N_{\rm vacua}\sim \frac{\pi^2}{108}\frac{(N_{\rm flux}^{\rm max})^2}{(t(m^2-4ln))^2},
\label{eq:Nvacua}
\end{align}
where $t(x)=x$ for $x\equiv 0$ $({\rm mod}\,3)$ and otherwise $t(x)=3x$, and $\{l,m,n\}$ determine the vacuum expectation value of the complex structure modulus $\tau$ by Eq. (\ref{eq:lnpositive}). 
Note that the analysis in Ref. \cite{DeWolfe:2004ns} adopts the parametrization ${\rm gcd}(l,m,n)=1$ in the equation $P(\tau)=0$ 
determining the vacuum expectation value of the modulus $\tau$, but the ratio between the number of vacua would be the same with our results when the continuous approximation of the flux quanta is valid, that is the $N_{\rm flux}^{\rm max}\gg 1$ regime. 
Hence, they would capture the phenomena of our results. 
The approximated probabilities of $\mathbb{Z}_3$, $\mathbb{Z}_2$ and other vacua shown in Tables 
\ref{tab:UNflux10} and \ref{tab:UNflux1000} are calculated as in Table \ref{tab:NvacuaProb}, in which 
we take the values of $(l,m,n)$ appearing in $P(\tau)=0$ such as $(l,m,n)=(1,0,1)$  for $\mathbb{Z}_2$ 
fixed point and $(l,m,n)=(1,-1,1)$ for $\mathbb{Z}_3$ fixed point, respectively. 
The number of total vacua was approximately calculated as $1.59\times 10^{-2}(N_{\rm flux}^{\rm max})^2$ \cite{DeWolfe:2004ns}. 
It turns out that $\mathbb{Z}_3$ fixed point is statistically favored in the string landscape. 
Furthermore, the denominator of Eq. (\ref{eq:Nvacua}) determines the magnitude of ${\rm Im}\,\tau$, 
which means that the smaller value of ${\rm Im}\,\tau$ is also favored in the string landscape. 
These phenomena are consistent with our results as seen in the right panel of Figure \ref{fig:1}. 

\begin{table}[H]
\scalebox{0.76}{
\begin{tabular}{|c|c|c|c|c|c|c|c|c|c|c|} \hline
$({\rm Re}\,\tau, {\rm Im}\,\tau)$ & ($-\frac{1}{2}, \frac{\sqrt{3}}{2}$) & ($0, \sqrt{3}$) & ($0, 1$) 
&($- \frac{1}{4}, \frac{\sqrt{15}}{4}$) & ($-\frac{1}{2}, \frac{\sqrt{15}}{2}$) & ($-\frac{1}{2}, \frac{\sqrt{7}}{2}$) & (0, $\sqrt{6}$) & ($0, \sqrt{\frac{3}{2}}$) & (0, $\sqrt{2}$) & ($-\frac{1}{2}, \frac{3\sqrt{3}}{2}$) \\ \hline
Probability ($\%$) & 63.9 & 3.99 & 3.99 & 2.56 & 2.56 & 1.30 & 1.00  & 1.00 & 1.00 & 0.79\\ \hline
\end{tabular}
}
\caption{Approximated probabilities of the stable vacua as functions of $({\rm Re}\,\tau, {\rm Im}\,\tau)$ shown 
in Table \ref{tab:UNflux1000}, predicted by the statistical approach.} 
\label{tab:NvacuaProb}
\end{table}

It is remarkable that the $\mathbb{Z}_3$ fixed point is favorable from the moduli stabilization.
At $\tau = \omega$, the $T^2/\mathbb{Z}_2$ orbifold has a high symmetry.
The four fixed points on $T^2/\mathbb{Z}_2$ are equally spaced like a tetrahedron.
The same stabilized value $\tau = \omega$ is realized by another scenario of the moduli stabilization \cite{Abe:2020vmv}.

From the phenomenological point of view, these results support the studies of modular flavor models around 
specific modulus points. 
Indeed, phenomenological aspects of modular flavor models are studied around the fixed points of the $SL(2,\mathbb{Z})$ 
modular group \cite{Novichkov:2018ovf,Novichkov:2018yse,Okada:2019uoy,Gui-JunDing:2019wap}.
The $\mathbb{Z}_3$ and $\mathbb{Z}_2$ residual symmetries remain on these fixed points.
Such symmetries control mass matrices of quarks and leptons.
Then, phenomenologically interesting results are obtained.
It is important that such fixed points are statically favored in the string landscape.
In addition to these fixed points, results in Tables \ref{tab:UNflux10} and \ref{tab:UNflux1000} suggest us to study other 
non-trivial moduli values with definite probabilities. 
It is interesting to explore the modular flavor models around these modulus values in addition to the fixed points.

So far, we have not imposed any bound for the axio-dilaton $S$. 
Let us next impose the bound for the axio-dilaton ${\rm Im}\,S$, focusing on the weak coupling 
regime. From the viewpoint of the string theory, the axio-dilaton determines the size of the string 
coupling, $g_s=({\rm Im}\,S)^{-1}$. Hence, the weak coupling regime is theoretically controllable against 
perturbative and non-perturbative corrections with respect to the string coupling. 
We calculate the probability of the stable vacua as functions of the complex structure modulus $\tau$ and 
the axio-dilaton $S$ as summarized in Tables \ref{tab:UwrtImS_Nflux10} and \ref{tab:UwrtImS_Nflux1000} 
for $N_{\rm flux}^{\rm max}=10$ and $N_{\rm flux}^{\rm max}=1000$, respectively. 
Here we list probabilities in the descending order of highest probabilities of $\tau$. 
It turns out that the distribution of the stable vacua for $N_{\rm flux}^{\rm max}=1000$ are uniformly distributed with respect 
to the axio-dilaton, 
but for the weak coupling as well as the small flux regime: ${\rm Im}\,S>1, N_{\rm flux}^{\rm max}=10$, 
all the stable vacua are clustered at the $\mathbb{Z}_3$ fixed point. 
In this respect, the theoretically controllable weak coupling region also prefers the $\mathbb{Z}_3$ fixed point in 
the fundamental domain of $SL(2,\mathbb{Z})$ moduli space. 

These results motivate us to study the contribution of the axio-dilaton to the distribution of the 
complex structure modulus.
That will be  discussed in the next section.

\begin{table}[htb]
\centering
\scalebox{0.85}{
\begin{tabular}{|c|c|c|c|c|c|c|c|} \hline
\diagbox[width=4cm]{(${\rm Re}\,\tau$, ${\rm Im}\,\tau$)}{${\rm Im}\,S$} & {\rm All} & $\frac{\sqrt{3}}{2}$ & 1 & 2
 & 3 & 4  & 5 \\ \hline
($-\frac{1}{2}, \frac{\sqrt{3}}{2}$) & 62.3 & 62.3 & 63.4 & 77.8 & 77.8 & 100 & 100 \\ \hline
($0, \sqrt{3}$) & 7.55 & 7.55 & 7.32 & 5.56 & 11.1 & 0 & 0  \\ \hline
 ($-\frac{1}{2}, \frac{\sqrt{15}}{2}$)  & 7.55 & 7.55 & 7.32 & 0 & 0 & 0 & 0 \\ \hline
($-\frac{1}{4}, \frac{\sqrt{15}}{4}$) & 7.55 & 7.55 & 4.88 & 5.56& 11.1 & 0 & 0 \\ \hline
($0, 1$) & 5.66 & 5.66 & 7.32 & 5.56 & 0 & 0 & 0 \\ \hline
($-\frac{1}{2}, \frac{\sqrt{7}}{2}$) & 1.89 & 1.89 & 2.44 & 0 & 0 & 0 & 0 \\ \hline
($0, \sqrt{6}$)  & 1.89 & 1.89 & 2.44 & 0 & 0 & 0 & 0 \\ \hline
($0, \sqrt{\frac{3}{2}}$) & 1.89 & 1.89 & 2.44 & 5.56 & 0 & 0 & 0 \\ \hline
(0, $\sqrt{2}$)  & 1.89 & 1.89 & 2.44 & 0 & 0 & 0 & 0 \\ \hline
($-\frac{1}{2}, \frac{3\sqrt{3}}{2}$) & 1.89 & 1.89 & 0 & 0 & 0 & 0 & 0 \\ \hline
\end{tabular}
}
\caption{Probabilities (\%) of the stable vacua as functions of $({\rm Re}\,\tau, {\rm Im}\,\tau)$ and ${\rm Im}\,S$ 
in the descending order of the probability with respect to $({\rm Re}\,\tau, {\rm Im}\,\tau)$. 
Here, we list the probability for $N_{\rm flux}^{\rm max}=10$ and the column in ``All'' corresponds 
to the probability of Table \ref{tab:UNflux10} .} 
\label{tab:UwrtImS_Nflux10}
\end{table}
\begin{table}[H]
\centering
\scalebox{0.85}{
\begin{tabular}{|c|c|c|c|c|c|c|c|} \hline
\diagbox[width=4cm]{(${\rm Re}\,\tau$, ${\rm Im}\,\tau$)}{${\rm Im}\,S$} & {\rm All} & $\frac{\sqrt{3}}{2}$ & 1 & 2
 & 3 & 4  & 5 \\ \hline
($-\frac{1}{2}, \frac{\sqrt{3}}{2}$) & 40.3 & 40.3 & 40.3 & 40.7 & 41.0 & 41.4 & 41.7 \\ \hline
($0, \sqrt{3}$) & 7.56 & 7.56 & 7.57 & 7.62 & 7.67 & 7.71 & 7.77  \\ \hline
 ($-\frac{1}{2}, \frac{\sqrt{15}}{2}$)  & 4.85 & 4.85 & 4.85 & 4.88 & 4.91 & 4.93 & 4.99 \\ \hline
($-\frac{1}{4}, \frac{\sqrt{15}}{4}$) & 4.85 & 4.85 & 4.85 & 4.88 & 4.92 & 4.92 & 4.98 \\ \hline
($0, 1$) & 3.79 & 3.79 & 3.82 & 3.85 & 3.86 & 3.89 & 3.90 \\ \hline
($-\frac{1}{2}, \frac{\sqrt{7}}{2}$) & 2.44 & 2.44 & 2.44 & 2.45 & 2.47 & 2.46 & 2.50 \\ \hline
($0, \sqrt{6}$) & 1.88 & 1.88 & 1.89 & 1.90 & 1.90 & 1.90 & 1.92 \\ \hline
($0, \sqrt{\frac{3}{2}}$) & 1.88 & 1.88 & 1.89 & 1.90 & 1.89 & 1.92 & 1.90 \\ \hline
(0, $\sqrt{2}$) & 1.88 & 1.88 & 1.89 & 1.90 & 1.91 & 1.92 & 1.91 \\ \hline
($-\frac{1}{2}, \frac{3\sqrt{3}}{2}$) & 1.49 & 1.49 & 1.49 & 1.50 & 1.51 & 1.52 & 1.49 \\ \hline
\end{tabular}
}
\caption{Probabilities (\%) of the stable vacua as functions of $({\rm Re}\,\tau, {\rm Im}\,\tau)$ and ${\rm Im}\,S$ 
in the descending order of the probability with respect to $({\rm Re}\,\tau, {\rm Im}\,\tau)$. 
Here, we list the probability for $N_{\rm flux}^{\rm max}=1000$ and the column in ``All'' corresponds 
to the probability of Table \ref{tab:UNflux1000} .} 
\label{tab:UwrtImS_Nflux1000}
\end{table}

\subsection{Distribution of complex structure modulus with the fixed axio-dilaton}
\label{subsec:3_2}

In this section, we further study the distribution of the complex structure modulus $\tau$, 
taking into account the distribution of the axio-dilaton $S$. 
As shown in Eq. (\ref{eq:Svev}), the vacuum expectation value of the axio-dilaton 
is correlated with that of the complex structure modulus $\tau$. 

\begin{figure}[htb]
\begin{minipage}{0.5\hsize}
  \begin{center}
   \includegraphics[height=120mm]{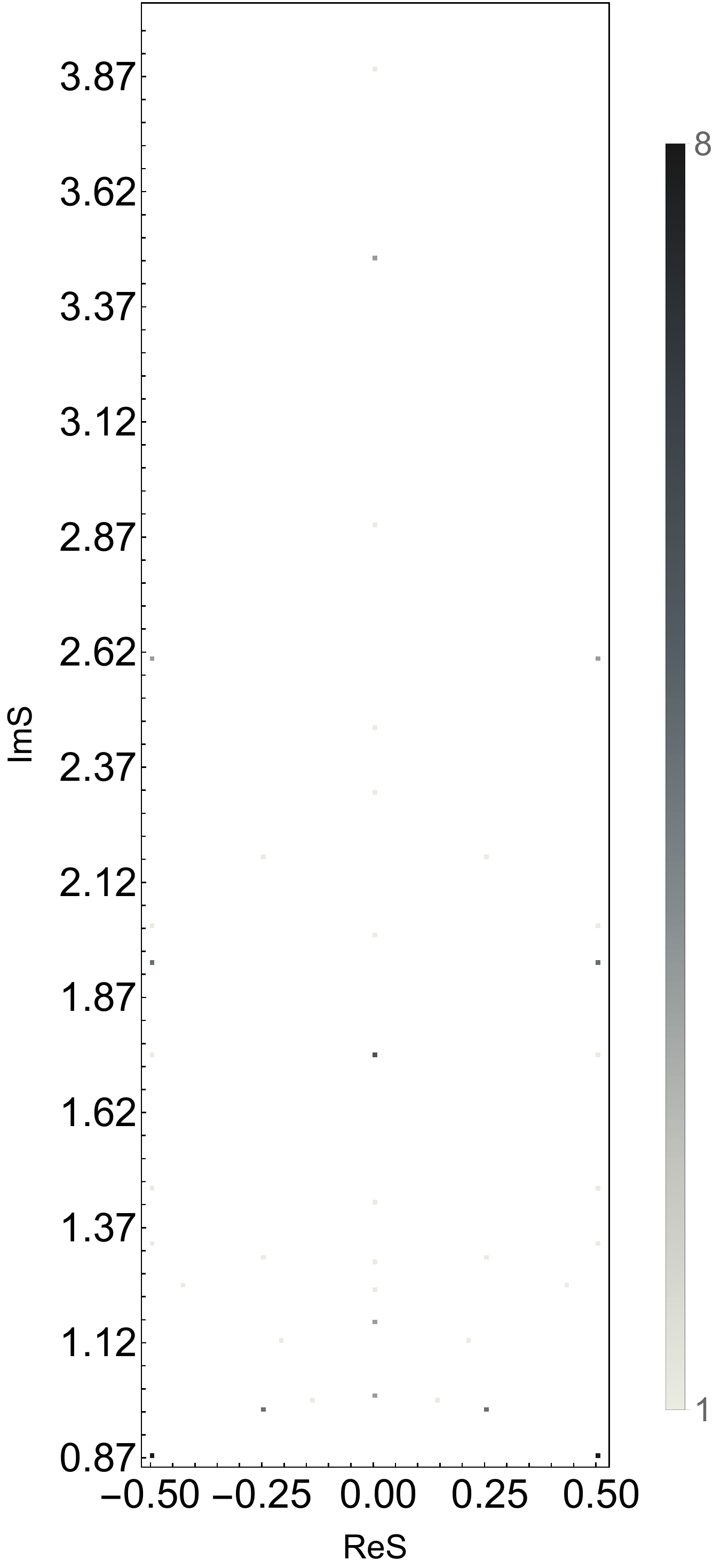}
  \end{center}
 \end{minipage}
 \begin{minipage}{0.5\hsize}
  \begin{center}
   \includegraphics[height=120mm]{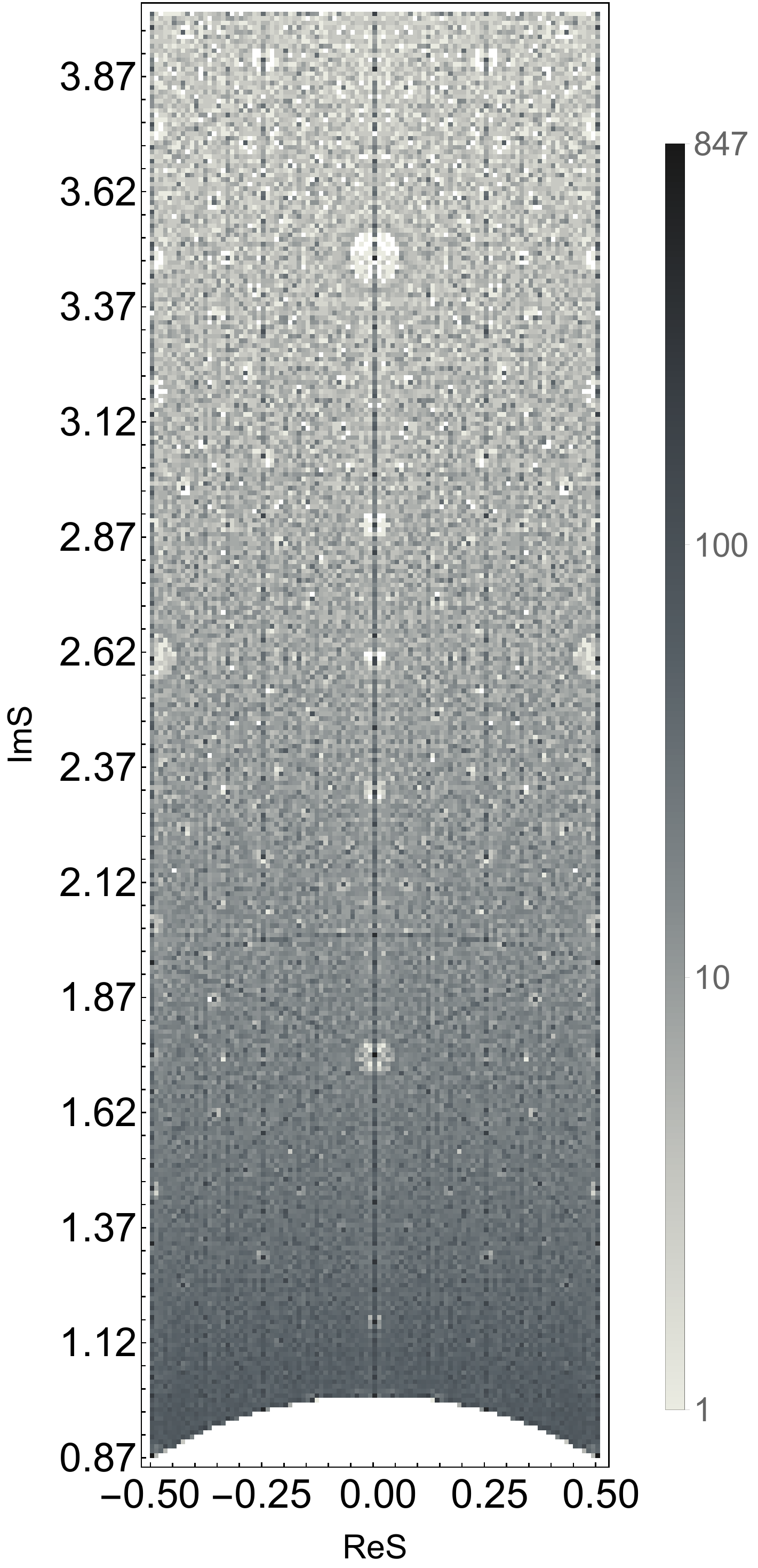}
  \end{center}
 \end{minipage}
  \caption{The numbers of the stable vacua on $({\rm Re}\,S, {\rm Im}\,S)$ plane for $N_{\rm flux}^{\rm max}=10$ 
  in the left panel and $N_{\rm flux}^{\rm max}=1000$ in the right panel, respectively.}
\label{fig:3_2}
\end{figure}

Let us work with the distribution of the axio-dilaton without imposing any bound for 
the complex structure modulus. 
We project out the distribution of the stable vacua on the plane of the axio-dilaton $S$ 
in the fundamental region of the $SL(2,\mathbb{Z})_S$ moduli space (\ref{eq:fund}), 
as drawn in Figure \ref{fig:3_2}, in which the stable vacua are clustered in the dark color region. 

There is a difference between the numbers of stable vacua for $N_{\rm flux}^{\rm max}=10$ and $N_{\rm flux}^{\rm max}=1000$ as shown in Table \ref{tab:Number}, but in both cases, the distributions of the stable vacua are not limited on 
the fixed points in the fundamental domain of $SL(2,\mathbb{Z})_S$ moduli space. 
Indeed, the probabilities of stable vacua in its descending order are summarized in Tables \ref{tab:SNflux10} 
and \ref{tab:SNflux1000}, from which the axio-dilaton is distributed along the strong coupling regime ${\rm Im}\,S\sim 1$ and there are no particular points on the plane of the axio-dilaton. 
From this point of view, we pick up 10 points in the descending order of the probability of the stable vacua on 
the moduli space of the axio-dilaton. 
Then, we discuss their consequences to the distributions of the complex structure modulus 
as summarized in Tables \ref{tab:Uwrt_specificImS_Nflux10} and \ref{tab:Uwrt_specificImS_Nflux1000}. 
It turns out that the residual $\mathbb{Z}_3$ fixed point is favoured in the landscape, but the 
existence of other points is sensitive to the value of the axio-dilaton. 
For example, $\mathbb{Z}_2$ fixed point $\tau=i$ can be realized on specific points in the moduli space of the 
axio-dilaton such as $S=i, 2i$, at which the $\mathbb{Z}_2$ fixed point is only allowed in the small $N_{\rm flux}^{\rm max}$ regime. Also, $\tau=\sqrt{2}i, \sqrt{3}i$ vacua are statistically favored on specific values of the axio-dilaton $S$. 

\begin{table}[H]
\scalebox{0.77}{
\begin{tabular}{|c|c|c|c|c|c|c|c|c|c|c|} \hline
$({\rm Re}\,S, {\rm Im}\,S)$ & ($-\frac{1}{2}, \frac{\sqrt{3}}{2}$) & ($0, \sqrt{3}$) & ($-\frac{1}{2}, \frac{\sqrt{15}}{2}$) 
 & ($-\frac{1}{4}, \frac{\sqrt{15}}{4}$) & ($-\frac{1}{2}, \frac{3\sqrt{3}}{2}$)  & (0,1) & ($0, 2\sqrt{3}$)  & ($0, \frac{2}{\sqrt{3}}$)  &  ($0, 5\sqrt{3}$) &($0, \frac{5}{\sqrt{3}}$) \\ \hline
Probability ($\%$) & 15.1 & 7.55 & 5.66 & 5.66 & 3.77 & 3.77 & 3.77 & 3.77 & 1.89 & 1.89 \\ \hline
\end{tabular}
}
\caption{Probabilities of the stable vacua as functions of $({\rm Re}\,S, {\rm Im}\,S)$ 
in the descending order of the probability for $N_{\rm flux}^{\rm max}=10$.} 
\label{tab:SNflux10}
\end{table}
\begin{table}[H]
\scalebox{0.77}{
\begin{tabular}{|c|c|c|c|c|c|c|c|c|c|c|} \hline
$({\rm Re}\,S, {\rm Im}\,S)$ & ($-\frac{1}{2}, \frac{\sqrt{3}}{2}$) & ($0, \sqrt{3}$) & ($-\frac{1}{2}, \frac{\sqrt{15}}{2}$) 
 & ($-\frac{1}{4}, \frac{\sqrt{15}}{4}$) & ($-\frac{1}{2}, \frac{\sqrt{7}}{2}$)  & ($0, 2\sqrt{3}$) & ($0, \frac{2}{\sqrt{3}}$)  & ($-\frac{1}{2}, \frac{3\sqrt{3}}{\sqrt{2}}$)  &  ($0, 1$) &($0, 2$) \\ \hline
Probability ($\%$) & 0.12 & 0.12 & 0.073 & 0.073 & 0.069 & 0.066 & 0.066 & 0.063 & 0.061 & 0.057 \\ \hline
\end{tabular}
}
\caption{Probabilities of the stable vacua as functions of $({\rm Re}\,S, {\rm Im}\,S)$ 
in the descending order of the probability for $N_{\rm flux}^{\rm max}=1000$.} 
\label{tab:SNflux1000}
\end{table}

\begin{table}[H]
\centering
\scalebox{0.7}{
\begin{tabular}{|c|c|c|c|c|c|c|c|c|c|c|c|} \hline
\diagbox[width=4cm]{(${\rm Re}\,\tau$, ${\rm Im}\,\tau$)}{(${\rm Re}\,S$, ${\rm Im}\,S$)} & {\rm All} & ($-\frac{1}{2}, \frac{\sqrt{3}}{2}$) & ($0, \sqrt{3}$) & ($-\frac{1}{2}, \frac{\sqrt{15}}{2}$) 
 & ($-\frac{1}{4}, \frac{\sqrt{15}}{4}$) & ($-\frac{1}{2}, \frac{\sqrt{7}}{2}$)  & ($0, 2\sqrt{3}$) & ($0, \frac{2}{\sqrt{3}}$)  & ($-\frac{1}{2}, \frac{3\sqrt{3}}{\sqrt{2}}$)  &  ($0, 1$) &($0, 2$) \\ \hline
($-\frac{1}{2}, \frac{\sqrt{3}}{2}$) & 62.3 & 75 & 75 & 0 & 0 & 0 & 50 & 50 & 100 & 0 & 0 \\ \hline
($0, \sqrt{3}$) & 7.55 & 12.5 & 25 & 0 & 0 & 0 & 50 & 50 & 0 & 0 & 0\\ \hline
 ($-\frac{1}{2}, \frac{\sqrt{15}}{2}$)  &7.55 & 0 & 0 & 66.7 & 33.3 & 0 & 0 & 0 & 0 & 0  & 0\\ \hline
($-\frac{1}{4}, \frac{\sqrt{15}}{4}$) & 7.55 & 0 & 0 & 33.3 & 66.7 & 0 & 0 & 0 & 0 & 0 & 0 \\ \hline
($0, 1$) & 5.66 & 0 & 0 & 0 & 0 & 0 & 0 & 0 & 0 &100 & 100\\ \hline
($-\frac{1}{2}, \frac{\sqrt{7}}{2}$) & 1.89& 0 & 0 & 0 & 0 & 100 & 0 & 0 & 0 & 0 & 0 \\ \hline
($0, 6$) & 1.89 & 0 & 0 & 0 & 0 & 0 & 0 & 0 & 0 & 0 & 0\\ \hline
($0, \sqrt{\frac{3}{2}}$) & 1.89 & 0 & 0 & 0 & 0 & 0 & 0 & 0 & 0 & 0 & 0 \\ \hline
(0, $\sqrt{2}$) & 1.89 & 0 & 0 & 0 & 0 & 0 & 0 & 0 & 0 & 0 & 0\\ \hline
($-\frac{1}{2}, \frac{3\sqrt{3}}{2}$) & 1.89 & 12.5 & 0 & 0 & 0 & 0 & 0 & 0 & 0 & 0 & 0\\ \hline
\end{tabular}
}
\caption{Probabilities of the stable vacua as functions of $({\rm Re}\,\tau, {\rm Im}\,\tau)$ and $({\rm Re}\,S, {\rm Im}\,S)$ 
in the descending order of the probability with respect to $({\rm Re}\,\tau, {\rm Im}\,\tau)$. 
Here, we list the probability for $N_{\rm flux}^{\rm max}=10$ and the column in ``All'' corresponds 
to the probability of Table \ref{tab:UNflux10} .} 
\label{tab:Uwrt_specificImS_Nflux10}
\end{table} 
\begin{table}[H]
\centering
\scalebox{0.7}{
\begin{tabular}{|c|c|c|c|c|c|c|c|c|c|c|c|} \hline
\diagbox[width=4cm]{(${\rm Re}\,\tau$, ${\rm Im}\,\tau$)}{(${\rm Re}\,S$, ${\rm Im}\,S$)} & {\rm All} & ($-\frac{1}{2}, \frac{\sqrt{3}}{2}$) & ($0, \sqrt{3}$) & ($-\frac{1}{2}, \frac{\sqrt{15}}{2}$) 
 & ($-\frac{1}{4}, \frac{\sqrt{15}}{4}$) & ($-\frac{1}{2}, \frac{\sqrt{7}}{2}$)  & ($0, 2\sqrt{3}$) & ($0, \frac{2}{\sqrt{3}}$)  & ($-\frac{1}{2}, \frac{3\sqrt{3}}{\sqrt{2}}$)  &  ($0, 1$) &($0, 2$) \\ \hline
($-\frac{1}{2}, \frac{\sqrt{3}}{2}$) & 40.3 & 73.0 & 36.8 & 0 & 0 & 0 & 33.6 & 33.6 & 100 & 0 & 0 \\ \hline
($0, \sqrt{3}$) & 7.56 & 9.01 & 27.5 & 0 & 0 & 0 & 25.4 &25.4 & 17.6 & 0 & 0\\ \hline
 ($-\frac{1}{2}, \frac{\sqrt{15}}{2}$)  &4.85 & 0 & 0 & 33.3 & 32.7 & 0 & 0 & 0 & 0 & 0  & 0\\ \hline
($-\frac{1}{4}, \frac{\sqrt{15}}{4}$) & 4.85 & 0 & 0 & 32.7 & 33.3 & 0 & 0 & 0 & 0 & 0 & 0 \\ \hline
($0, 1$) & 3.79 & 0 & 0 & 0 & 0 & 0 & 0 & 0 & 0 & 47.6 & 25.8 \\ \hline
($-\frac{1}{2}, \frac{\sqrt{7}}{2}$) & 2.44& 0 & 0 & 0 & 0 & 35.7 & 0 & 0 & 0 & 0 & 0 \\ \hline
($0, 6$) & 1.88 & 0 & 0 & 0 & 0 & 0 & 0 & 0 & 0 & 0 & 0\\ \hline
($0, \sqrt{\frac{3}{2}}$) & 1.88 & 0 & 0 & 0 & 0 & 0 & 0 & 0 & 0 & 0 & 0 \\ \hline
(0, $\sqrt{2}$) & 1.88 & 0 & 0 & 0 & 0 & 0 & 0 & 0 & 0 & 0 & 0\\ \hline
($-\frac{1}{2}, \frac{3\sqrt{3}}{2}$) & 1.88 & 8.17 & 12.1 & 0 & 0 & 0 & 10.6 &  10.6 & 16.2& 0 & 0\\ \hline
\end{tabular}
}
\caption{Probabilities of the stable vacua as functions of $({\rm Re}\,\tau, {\rm Im}\,\tau)$ and $({\rm Re}\,S, {\rm Im}\,S)$ 
in the descending order of the probability with respect to $({\rm Re}\,\tau, {\rm Im}\,\tau)$. 
Here, we list the probability for $N_{\rm flux}^{\rm max}=1000$ and the column in ``All'' corresponds 
to the probability of Table \ref{tab:UNflux1000} .} 
\label{tab:Uwrt_specificImS_Nflux1000}
\end{table}

\subsection{Distributions of CP-breaking vacua}
\label{subsec:3_3}

Finally, we study the distribution of CP-breaking vacua. 
It is known that the 4D CP symmetry is embedded into the 10D proper Lorentz symmetry 
in terms of the 4D parity and the 6D orientation reversing transformations \cite{Strominger:1985it}. 
Here, we assume that the 10D spacetime is given by the product of 4D spacetime and 
6D Calabi-Yau manifolds including the torus. 
Since the 6D orientation reversing transformation changes the sign of the 6D volume form $dV$, 
it corresponds to the anti-holomorphic involution of the holomorphic three-form of 6D space. 
Given the holomorphic three-form $\Omega=dz_1\wedge dz_2\wedge dz_3$ in the local coordinates 
of the 6D space $\{z_1,z_2,z_3\}$, the 6D orientation reversing transformation 
corresponds to
\begin{align}
    \Omega \xrightarrow{{\rm CP}} -\bar{\Omega},
\end{align}
due to the property $i\Omega \wedge \bar{\Omega}\propto dV$. (See for more details in 
Ref. \cite{Ishiguro:2020nuf}.) 
In the factorizable torus $T^6$ and its orbifolds, the 6D orientation reversing 
transformation is provided by the anti-holomoprhic involution of the complex structure,
\begin{align}
    \tau_i \xrightarrow{{\rm CP}} -\bar{\tau_i}.
\label{eq:tauCPtrf}
\end{align}
Hence, the 4D CP symmetry is unbroken at ${\rm Re}\,\tau_i=0$. 
Furthermore, there exist the additional CP-conserving region in the fundamental region of the moduli space 
in models with modular symmetry \cite{Novichkov:2019sqv}\footnote{In Ref.~\cite{Kobayashi:2019uyt}, it was shown that 
the CP phase can be rephased concretely in modular flavor models.}. 

As the simplest example having the CP symmetry, ${\rm Re}\,\tau_i=\pm 1/2$ are the CP-conserving lines, 
since they are invariant under Eq.~(\ref{eq:tauCPtrf}) up to the $T$-transformation, namely
\begin{align}
    {\rm Re}\,\tau_i =\pm \frac{1}{2} \xrightarrow{{\rm CP}}\mp \frac{1}{2} \xrightarrow{T} 
    {\rm Re}\,\tau_i =\pm \frac{1}{2}.
\end{align}
In addition to it, the locus $|\tau_i|=1$ is also CP-invariant up to the $S$-transformation. 
On this locus, the $S$-transformation is identified with the CP-transformation; 
\begin{align}
    \tau_i \xrightarrow{S} -\frac{1}{\tau_i} = -\frac{\bar{\tau}_i}{|\tau_i|}= -\bar{\tau}_i.
\end{align}
Thus, the CP symmetry remains on ${\rm Re}\,\tau_i=0$ and the boundary 
of the $SL(2,\mathbb{Z})$ fundamental domain, that is, the boundary of the $GL(2,\mathbb{Z})$ fundamental region.
This argument is applicable to the moduli space of the axio-dilaton. 
So far, we have considered the factorizable 6-torus with different complex structure moduli, but 
the above property also holds for the isotropic torus $\tau=\tau_1=\tau_2=\tau_3$. 
The breakdown of the CP symmetry is of particular importance to determine the size of 
Cabbibo-Kobayashi-Maskawa and Maki-Nakagawa-Sakata phases through the moduli-dependent Yukawa couplings. 
Indeed, in both bottom-up and top-down approaches (such as magnetized D-branes in Type IIB string 
setup), Yukawa couplings depend on the complex structure moduli $\tau$.

\begin{figure}[H]
\begin{minipage}{0.5\hsize}
  \begin{center}
   \includegraphics[height=74mm]{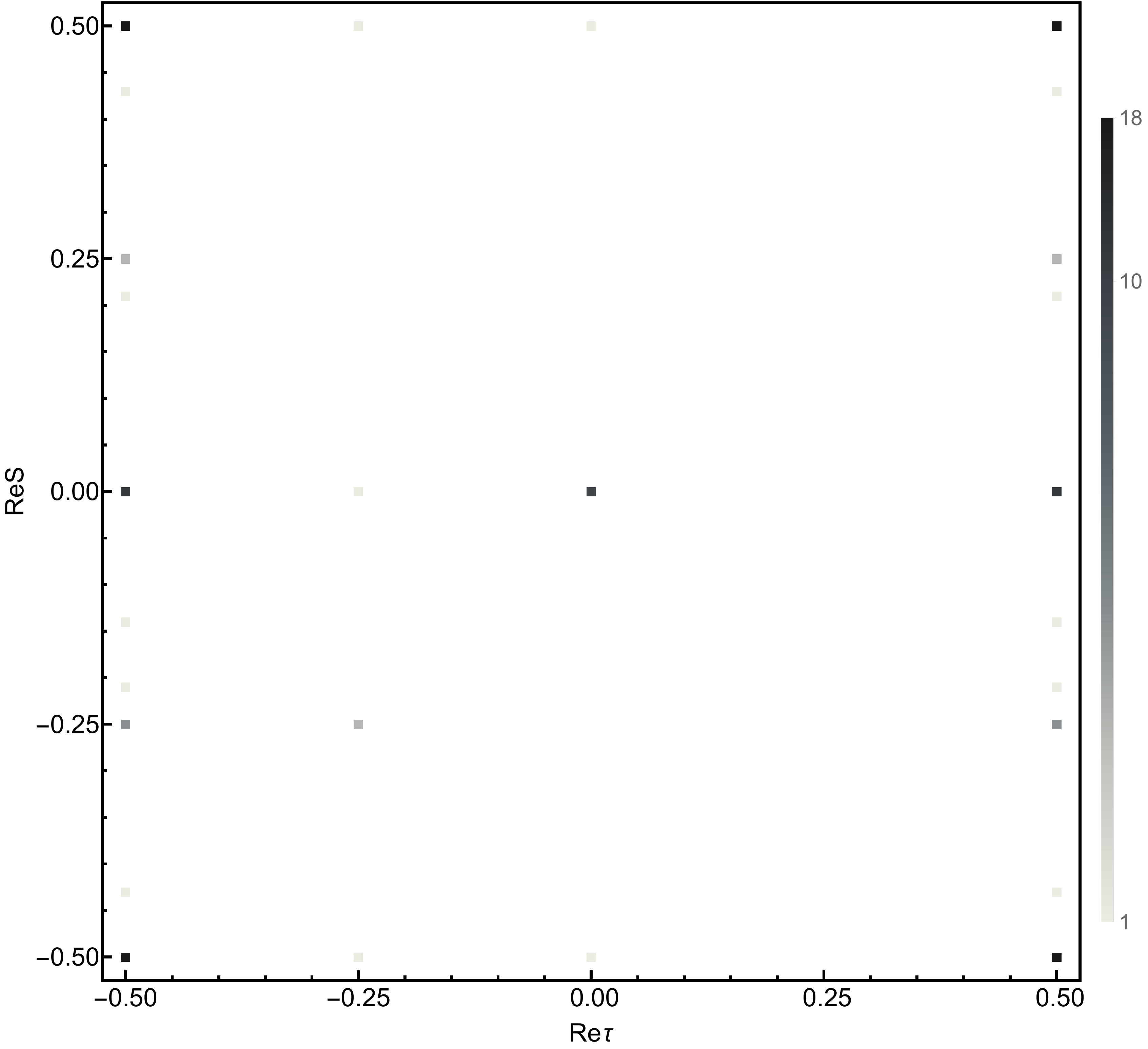}
  \end{center}
 \end{minipage}
 \begin{minipage}{0.5\hsize}
  \begin{center}
   \includegraphics[height=74mm]{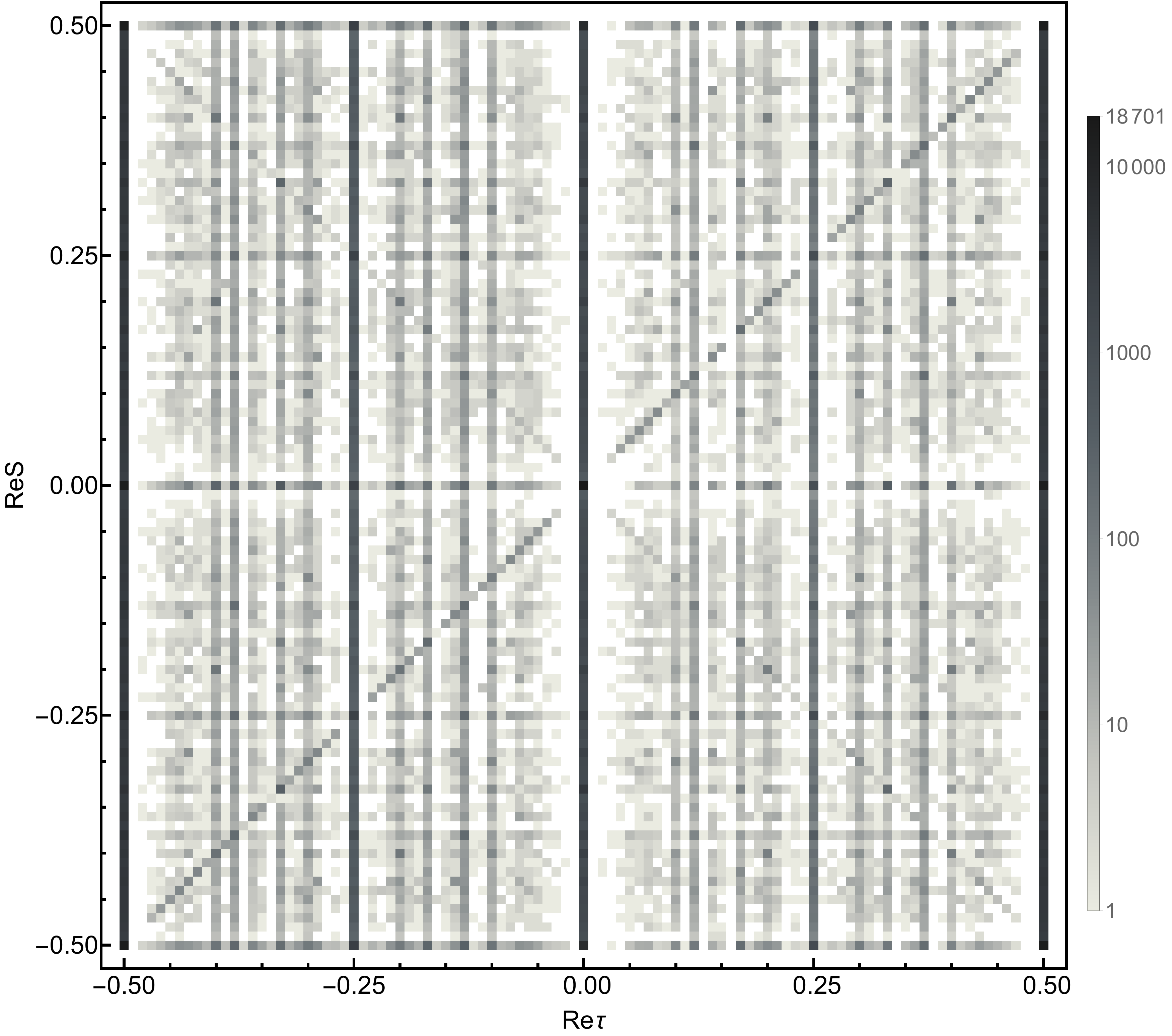}
  \end{center}
 \end{minipage}
    \caption{The numbers of stable vacua on the $({\rm Re}\,\tau,{\rm Re}\,S)$ plane for $N_{\rm flux}^{\rm max}=10$ 
    in the left panel and $N_{\rm flux}^{\rm max}=1000$ in the right panel, respectively.}
    \label{fig:ReSReU}
\end{figure}

We examine the distributions of CP-breaking supersymmetric vacua, allowing for general RR and NS fluxes which generically break the CP invariance of the action. 
Hence, it is not a scenario of spontaneous CP violation, but the CP symmetry would be spontaneously broken 
by particular choices of fluxes. 
Given the analytical solution satisfying the tadpole cancellation condition (\ref{eq:Nflux}), 
we find that the axionic fields $({\rm Re}\,\tau,{\rm Re}\,S)$ are mostly stabilized at the line ${\rm Re}\,\tau=0$ or the boundary of the $SL(2,\mathbb{Z})$ moduli spaces, especially in the small $N^{\rm max}_{\rm flux}$ case as shown in Figure \ref{fig:ReSReU}. 
The probabilities of the stable vacua as functions of $({\rm Re}\,\tau,{\rm Re}\,S)$ are summarized in 
Table \ref{tab:ReUReSNflux10} for $N_{\rm flux}^{\rm max}=10$ and Table \ref{tab:ReUReSNflux1000} for $N_{\rm flux}^{\rm max}=1000$, respectively. 
Hence, we conclude that CP-breaking vacua are disfavored in the complex structure moduli sector. 
This result is in agreement with the statistical approach. 
From the number of vacua as function of the vacuum expectation value of $\tau$ in Eq. (\ref{eq:Nvacua}), 
$m=0$ vacua are statistically favored due to the suppression of the denominator. 
Since the $m=0$ vacua correspond to ${\rm Re}\,\tau=0$ in Eqs. (\ref{eq:lnpositive}) and (\ref{eq:lnnegative}), 
the CP-conversing vacua are favored in the context of the statistical approach. 
Furthermore,  ${\rm Re}\,S$ seems to be more likely to lead CP-breaking vacua than ${\rm Re}\,\tau$, indicating from Table \ref{tab:ReUReS_CPbreakingratio}. 

On the other hand, the CP-breaking vacua in Eq.(\ref{eq:CPV-min}) have ${\cal O}(1)\%$ of probabilities.
In particular, when ${\rm Re}\,S=\pm 1/4$, the CP-breaking vacua for the complex structure modulus, 
\begin{align}
{\rm Re}\,\tau =\pm \frac{1}{4}
\label{eq:Retau14}
\end{align}
are statistically favored in the landscape.
This correlation may be important from the viewpoint of CP-violation phenomena.
Also, Figure~\ref{fig:ReSReU} shows the linear correlations, i.e., ${\rm Re}\,\tau = \pm {\rm Re }\,S$.
These results suggest that different CP-violating phases such as CP phases in mass matrices and the 
strong CP phase may have a common origin \footnote{See for a relevant scenario, Ref.~\cite{Kobayashi:2020oji}.}. 
We comment on the reason why the vacua in Eq. (\ref{eq:Retau14}) is 
favored in the complex structure modulus from the viewpoint of 
the statistical approach. 
We come back to the denominator in Eq. (\ref{eq:Nvacua}), which 
determines the statistically favored moduli values. 
This denominator is rewritten in terms of the 
vacuum expectation value of the modulus field (\ref{eq:lnpositive}). 
\begin{align}
t(m^2 -4ln)\propto
  -4l^2 ({\rm Im}\,\tau)^2, 
\end{align}
which indicates that the small $l$ enhances the number of vacua, in particular for 
$l=\pm 1$. 
In that $l=\pm 1$ case, together with the relation
\begin{align}
    m = -2l\,{\rm Re}\,\tau,
\end{align}
the integer $m$ also requires 
\begin{align}
    {\rm Re}\,\tau =\pm \frac{1}{4}.
\end{align}
Hence, the existence of CP-breaking vacua ${\rm Re}\,\tau=\pm 1/4$ are also 
favored in the statistical approach.

\begin{table}[H]
\scalebox{0.75}{
\begin{tabular}{|c|c|c|c|c|c|c|c|c|c|c|} \hline
$({\rm Re}\,\tau, {\rm Re}\,S)$ & ($-\frac{1}{2},-\frac{1}{2}$) &($-\frac{1}{2}, 0 $)   &($0,0$)  & ($-\frac{1}{2}, - \frac{1}{4}$) & ($-\frac{1}{2}, \frac{1}{4}$) &($- \frac{1}{4}, - \frac{1}{4}$) & ($0, - \frac{1}{2}$)  & ($ - \frac{1}{2}, \frac{3}{7}$) &($  - \frac{1}{2}, -\frac{3}{7}$) & ($-\frac{1}{2}, -\frac{3}{14}$)\\ \hline
Probability ($\%$) & 34.0 & 20.8 & 17.0 & 5.66 & 3.77 & 3.77 & 1.89 & 1.89 & 1.89 & 1.89 \\ \hline
\end{tabular}
}
\caption{Probabilities of the stable vacua as functions of $({\rm Re}\,\tau, {\rm Re}\,S)$ 
in the descending order of the probability for $N_{\rm flux}^{\rm max}=10$.} 
\label{tab:ReUReSNflux10}
\end{table}
\begin{table}[H]
\scalebox{0.79}{
\begin{tabular}{|c|c|c|c|c|c|c|c|c|c|c|} \hline
$({\rm Re}\,\tau, {\rm Re}\,S)$ & ($0,0$) & ($-\frac{1}{2},-\frac{1}{2}$) & ($-\frac{1}{2}, 0$)  & ($0, -\frac{1}{2}$) & ($-\frac{1}{2}, - \frac{1}{4}$) & ($-\frac{1}{2}, \frac{1}{4}$)  & ($0, - \frac{1}{3}$)  &  ($0, \frac{1}{3}$) &($-\frac{1}{4}, -\frac{1}{2}$) & ($0, -\frac{1}{4}$)\\ \hline
Probability ($\%$) & 2.74 & 2.18 & 1.98 & 1.02 & 0.830 & 0.801 & 0.468 & 0.447 & 0.380 & 0.355 \\ \hline
\end{tabular}
}
\caption{Probabilities of the stable vacua as functions of $({\rm Re}\,\tau, {\rm Re}\,S)$ 
in the descending order of the probability for $N_{\rm flux}^{\rm max}=1000$.} 
\label{tab:ReUReSNflux1000}
\end{table}
\begin{table}[H]
\centering
\scalebox{1.00}{
\begin{tabular}{|c|c|c|c|c|c|} \hline
 $N^{\rm max}_{\rm flux}$ & 10 & 100 & 250 & 500 & 1000\\ \hline
 ${\rm Re}\,S$& 22.6 & 66.3 & 78.7 & 85.3 & 90.0 \\ \hline
 ${\rm Re}\,\tau$ & 7.55 & 12.4 & 15.0 & 16.5 & 17.6\\ \hline
\end{tabular}
}
\caption{Probabilities (\%) of the CP-breaking vacuum expectation value of the axionic fields for various $N^{\rm max}_{\rm flux}$. } 
\label{tab:ReUReS_CPbreakingratio}
\end{table}

\section{Modular symmetric flavor models}
\label{sec:4}

In this section, to illustrate implications of our results, 
we compare the predictions in modular $A_4$ models 
with the distributions of complex structure modulus $\tau$ in flux vacua 
as shown in Section \ref{sec:3}. 
In particular, we focus on 8 classes of models in Ref. \cite{Ding:2019zxk} 
as summarized in Section \ref{subsec:4_1}. 
Section \ref{subsec:4_2} is devoted to the distributions of 
the complex structure modulus $\tau$ without fixing the axio-dilaton $S$. 
Furthermore, these distributions predict favorable vacuum expectation values 
of the moduli fields consistent with the observational data in 8 classes of modular $A_4$ models.

\subsection{Modular $A_4$ models}
\label{subsec:4_1}

We briefly review modular $A_4$ models explaining  the flavor structure. 
Before going to the detail of these models, we recall that there exist finite non-Abelian groups in the $SL(2,\mathbb{Z})$ modular group. 
The principal congruence subgroup $\Gamma(N)$ of $\Gamma=SL(2,\mathbb{Z})$ and its quotient groups are summarize as follows:
\begin{align}
    \Gamma(N) &=\left\{
    \biggl(
    \begin{array}{cc}
        p & q \\
        s & t
    \end{array}
    \biggl)
    \in \Gamma
    \biggl|
    p=t\equiv 1,\quad q=s\equiv 0\,({\rm mod}N)
    \right\},
    \nonumber\\
    \Gamma_N^\prime &\equiv \Gamma/\Gamma(N)
    =\langle S, T| S^4=(ST)^3=T^N=\mathbb{I},S^2T=TS^2\rangle.
\end{align}
Since the complex structure is invariant under the transformation $S^2$, 
the complex structure moduli space is governed by $\bar{\Gamma}\equiv \Gamma/\{\pm \mathbb{I}\}$. 
Its quotient groups are also defined as
\begin{align}
    \bar{\Gamma}(N)&\equiv \Gamma(N)/\{\pm \mathbb{I}\},
    \nonumber\\
    \Gamma_N&\equiv \bar{\Gamma}/\bar{\Gamma}(N)
    =\langle S^2=(ST)^3=T^N=\mathbb{I}\rangle,
\end{align}
and $\Gamma_{2,3,4,5}$ are isomorphic to the phenomenologically attractive $S_3, A_4, S_4, A_5$ non-Abelian discrete groups, respectively. 

To discuss the Yukawa couplings of quarks and leptons, modular form $f(\tau)$ of weight $k$ 
and level $N$ is of particular importance to 
be defined to satisfy
\begin{align}
    f\left(\frac{p\tau+q}{s\tau +t} \right)= (s\tau+t)^k f(\tau),
\end{align}
where $p,q,s,t$ denote the element of $\Gamma(N)$. In particular, one can assign the standard model particles to be irreducible representations of the modular group $\Gamma_N$. 
When we denote $F(\tau)=(f_1(\tau),f_2(\tau),\cdots)^T$ represent multiplets of modular forms transforming in the irreducible representations of $\Gamma_N$, 
it obeys
\begin{align}
    F(\gamma\tau)=(s\tau+t)^k\rho(\gamma)F(\tau),
\end{align}
for $\gamma\in \bar{\Gamma}(1)$, 
where $\rho(\tau)$ denotes a certain unitary representation matrix.

In this paper, we focus on the $\Gamma_3 \simeq A_4$ modular group which corresponds to a minimal subgroup 
admitting a triplet representation as well as three singlets. 
The singlet representations have the following representations under $S$- and $T$-transformations:
\begin{align}
    \mathbf{1}: (S=1,T=1),\quad
    \mathbf{1}^\prime: (S=1,T=w^2),\quad
    \mathbf{1}^{\prime\prime}: (S=1,T=w),\quad
\end{align}
whereas the triplet representation is assigned to be transformed under $S$- and $T$-transformations:
\begin{align}
    S=\frac{1}{3}
    \begin{pmatrix}
     -1 & 2 & 2 \\
     2 & -1 & 2 \\
     2 & 2 & -1\\
    \end{pmatrix}
    ,\quad
    T=
    \begin{pmatrix}
     1 & 0 & 0 \\
     0 & w^2 & 0 \\
     0 & 0 & w\\
    \end{pmatrix}
.
\end{align}
Since the modular forms of the weight 2 are spanned on three-dimensional modular space, 
these modular forms of the weight 2 are assigned to the triplet of $A_4$. Explicit forms of modular forms of the weight 2 are 
constructed in Ref.~\cite{Feruglio:2017spp}, and higher weight modular forms such as weights 4 and 6 are also constructed by tensor products of the modular forms of weight 2. 

We follow the supersymmetric modular $A_4$ models proposed in 
Ref. \cite{Ding:2019zxk}, in which comprehensive 
analysis of the lepton sector is performed 
under the smallest number of free parameters. 
In this analysis, we assign the charge assignment of $SU(2)$, $A_4$ and the modular weight for 
left-handed lepton doublets $L=(L_1,L_2,L_3)^T$, right-handed neutrino $N^c=(N_1^c,N_2^c,N_3^c)^T$,  the right-handed charged leptons $E^c=(E_1^c,E_2^c,E_3^c)$ and 
the Higgs doublets $H_u, H_d$ as shown in Table \ref{tab:charge}. 
\begin{table}[htb]
\centering
\scalebox{1}{
\begin{tabular}{|c||c|c|c|c|c|} \hline
& L & $N^c$ & $E_1^c$, $E_2^c$, $E_3^c$ & $H_u$ & $H_d$ \\ \hline \hline
$SU(2)$ & 2 & 1 & 1 & 2 & 2 \\
$A_4$ & $\mathbf{3}$ & $\mathbf{3}$ &all combinations of $\mathbf{1}$,$\mathbf{1}^{\prime}$, $\mathbf{1}^{\prime\prime}$  & $\mathbf{1}$ & $\mathbf{1}$ \\
$-k_I$ & $-k_L$ & $-k_{N^c}$  & $-k_{E_1}$,$-k_{E_2}$,$-k_{E_3}$ & 0 & 0 \\ \hline
\end{tabular}
}
\caption{Charge assignment of the lepton sector and Higgs doublets.} 
\label{tab:charge}
\end{table}

The modular weights of those fields, $-k_I$, are 
free parameters, but they are constrained to obtain the modular-invariant superpotential. 
Since the Yukawa couplings transform as the 
unitary representation $\rho_Y$ of the modular group $\Gamma_N$, the superpotential
\begin{align}
    W= \sum_n Y_{i_1\cdots i_n} \Phi_{i_1}\cdots \Phi_{i_n},
\end{align}
is modular invariant only if the modular weight of the coupling $Y_{i_1\cdots i_n}$, $k_Y$, satisfy 
$k_Y=\sum_n k_{i_n}$ (in global supersymmetric theory) with $k_{i_n}$ being the modular weight of fields $\Phi_{i_n}$. 
Furthermore, the product of representation $\rho_Y(\gamma)$ of the Yukawa coupling and 
that of fields should include the singlet representation, namely $\rho_Y \otimes_n \rho_{i_n} \ni \mathbf{1}$. 

To realize the tiny neutrino masses, the authors of Ref. \cite{Ding:2019zxk} discussed two scenarios. 
First one is to introduce the Weinberg operator
$W = \frac{1}{\Lambda}(H_uH_uLLY)_{\mathbf{1}}$ 
with cutoff scale $\Lambda$, leading to ten models, ${\cal A}_i$ ($i=1,\cdots, 10$)
and the other is to consider the type-I seesaw 
mechanism, leading to 30 models ${\cal B}_i, {\cal C}_i, {\cal D}_i$ ($i=1,\cdots, 10$). 
From the totally 40 classes of these models, 
it was found that 14 models are fitted with the 
observational data for both normal ordering and inverted ordering neutrino mass spectra. 
We compare these phenomenological results 
with the distribution of the complex structure 
modulus $\tau$ predicted from the string landscape in Section \ref{sec:3}.

\subsection{Distribution of complex structure modulus in modular $A_4$ models}
\label{subsec:4_2}

We are now ready to examine
predictions of modular $A_4$ models in light of our results on the moduli stabilization.
As for the modular $A_4$ models, we focus on the successful 8 models labeled by 
${\cal B}_9, {\cal B}_{10}, {\cal D}_{5,6,7,8,9,10}$, in which the type-I seesaw mechanism are employed to realize the tiny neutrino masses with the normal ordering. 
In these models, 6 (real) free parameters and two overall mass scales successfully lead to explain the flavor structures of the lepton sector, that is, the mass hierarchy of charged lepton masses, differences of neutrino masses squared and mixing angles.

\begin{table}[htb]
\centering
\scalebox{1}{
\begin{tabular}{|c||c|c|} \hline
 & \multicolumn{2}{c|}{Allowed regions}\\ \cline{2-3}
Models & for Re$\tau$ & for Im$\tau$\\
\hline 
\hline
${\rm Model}\, {\cal B}_9$  &  
[0, 0.368] 
&
[1.351, 1.856] 

\\ \hline
${\rm Model}\, {\cal B}_{10}$   &  
[0, 0.431] 
&
$[0.905, 1.168] \bigcup [1.305, 1.861]$

\\ \hline
${\rm Model}\, {\cal D}_5$  &  
[0.248, 0.300] 
&
[0.957, 1.056] 

\\ \hline
${\rm Model}\, {\cal D}_6$  &  
[0, 0.300] 
&
[0.957, 1.394] 
\\ \hline
${\rm Model}\, {\cal D}_7$  &  
[0.026, 0.5] 
&
[1.468, 3.006] 

\\ \hline
${\rm Model}\,{\cal D}_8$  &  
[0.424, 0.5] 
&
[0.872, 0.964] 

\\ \hline
${\rm Model}\ {\cal D}_9$  &  
$[0.033, 0.056] \bigcup [0.440, 0.469]$ 
&
$[0.887, 0.908] \bigcup [2.0, 2.282]$ 

\\ \hline
${\rm Model}\, {\cal D}_{10}$  &  
[0.0307, 0.1175] 
&
[1.996, 2.50] 
\\ \hline
\end{tabular}
}
\caption{Phenomenologically allowed regions of $({\rm Re}\,\tau, {\rm Im}\,\tau)$ \cite{Ding:2019zxk}.} 
\label{tab:A4models}
\end{table}

The allowed regions for $({\rm Re}\,\tau, {\rm Im}\,\tau)$ in 8 models are summarized in Table \ref{tab:A4models}, which provide the constrained observables such as the mixing angles of the lepton sector, 
relatively large neutrino masses and maximal Dirac CP phase. 
Note that the analysis in Ref.~\cite{Ding:2019zxk} focuses on the following moduli space:
\begin{align}
{\rm Re}\,\tau>0,\quad |\tau|>1.
\end{align}
That means that the $\mathbb{Z}_2$ fixed point $({\rm Re}\,\tau,{\rm Im}\,\tau)=(0,1)$, 
the $\mathbb{Z}_3$ fixed point $({\rm Re}\,\tau,{\rm Im}\,\tau)=(-\frac{1}{2},\frac{\sqrt{3}}{2})$ and 
$({\rm Re}\,\tau,{\rm Im}\,\tau)=(-\frac{1}{4},\frac{\sqrt{15}}{4})$ point are not taken into account 
in the analysis of Ref.~\cite{Ding:2019zxk}, although these points lead to high probabilities in our results from the string landscape. 
The allowed regions of ${\cal D}_8$, (${\cal B}_{10},{\cal D}_6$) and (${\cal B}_{10},{\cal D}_5$) models 
have values close to the $\mathbb{Z}_2$, $\mathbb{Z}_3$ and $({\rm Re}\,\tau,{\rm Im}\,\tau)=(-\frac{1}{4},\frac{\sqrt{15}}{4})$ points, respectively. 

\begin{figure}[H]
\begin{minipage}{0.5\hsize}
  \begin{center}
   \includegraphics[width=50mm]{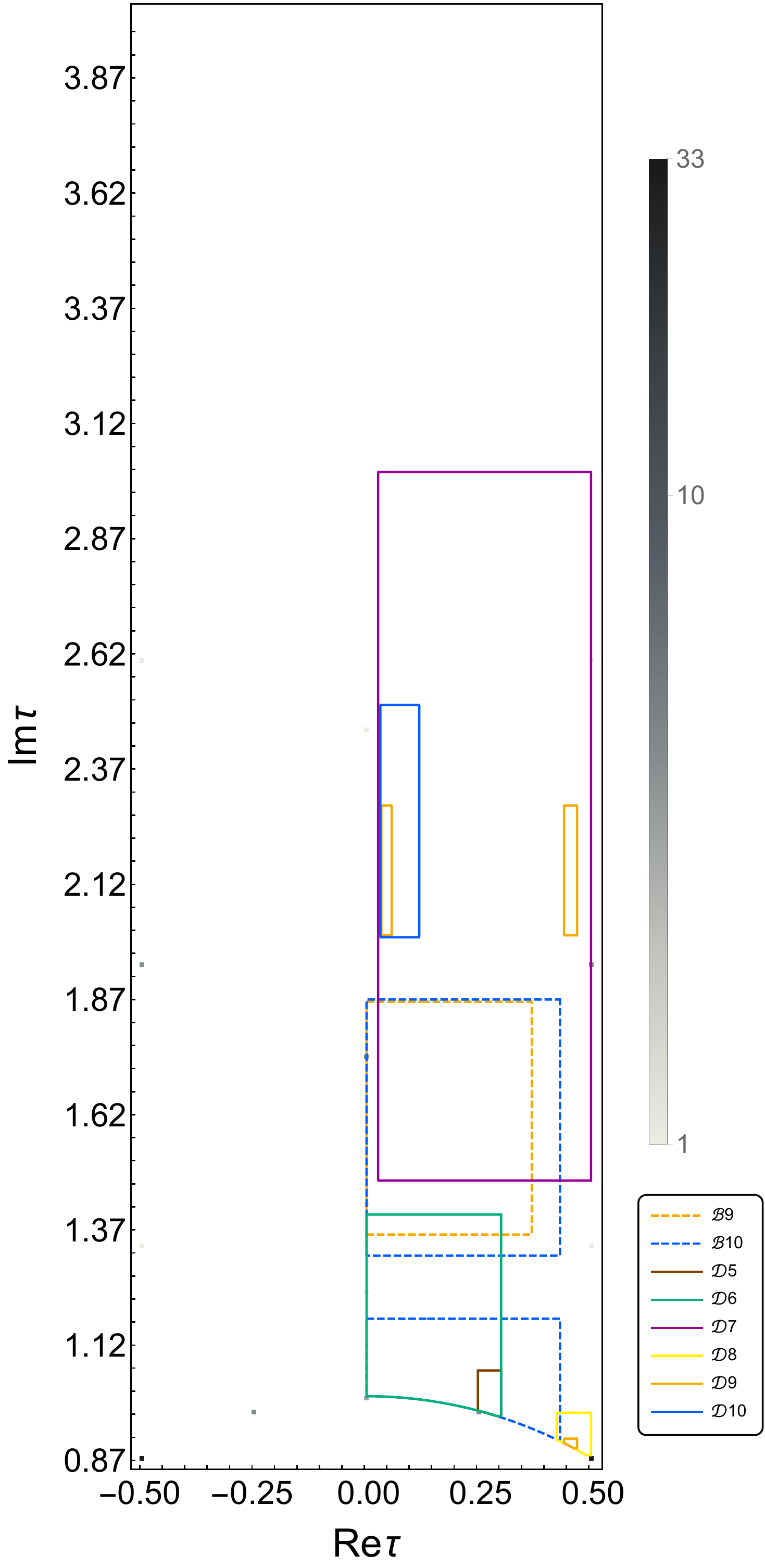}
  \end{center}
 \end{minipage}
 \begin{minipage}{0.5\hsize}
  \begin{center}
   \includegraphics[width=50mm]{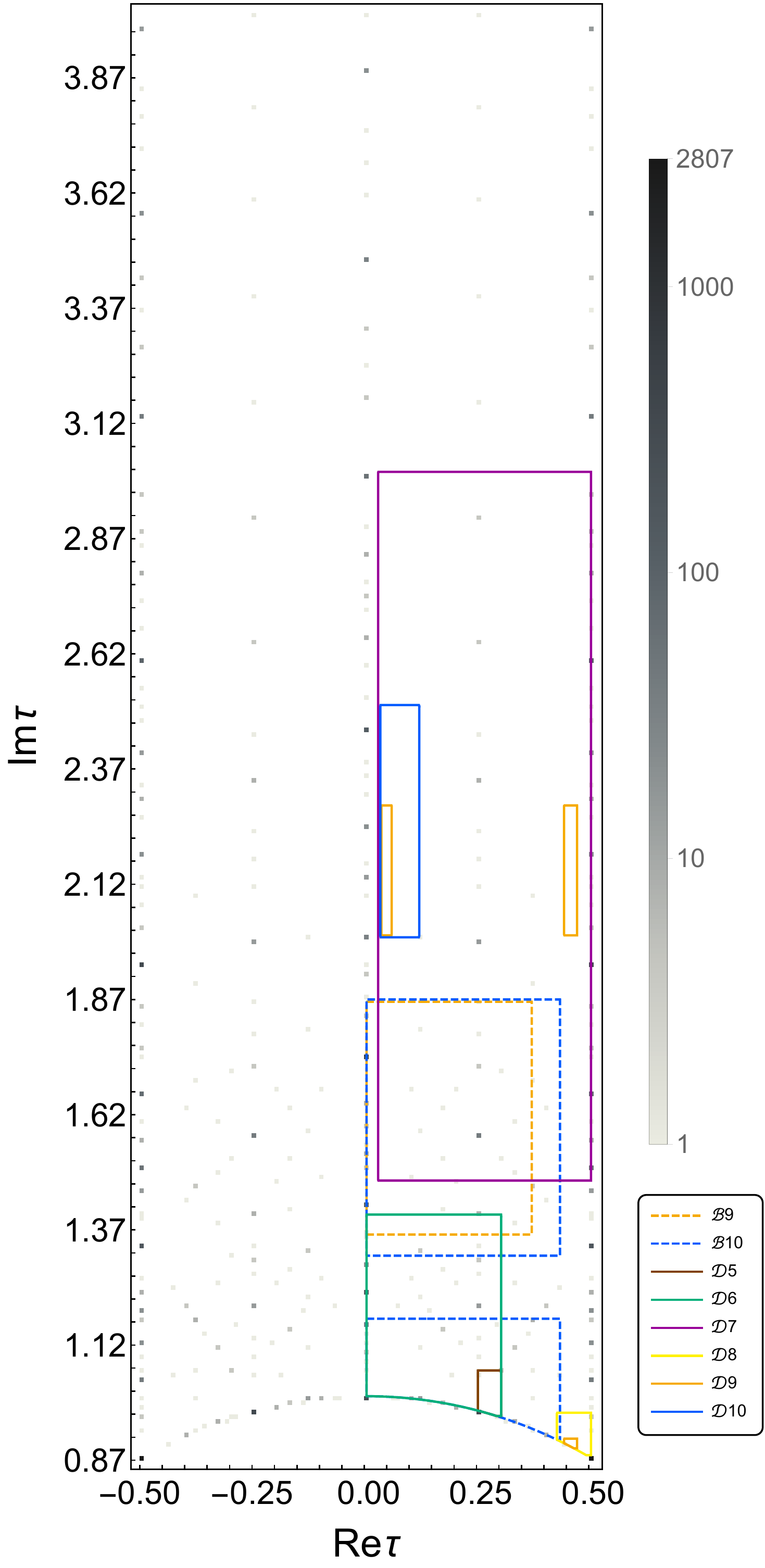}
  \end{center}
 \end{minipage}
 \begin{minipage}{0.507\hsize}
  \begin{center}
   \includegraphics[width=51.7mm]{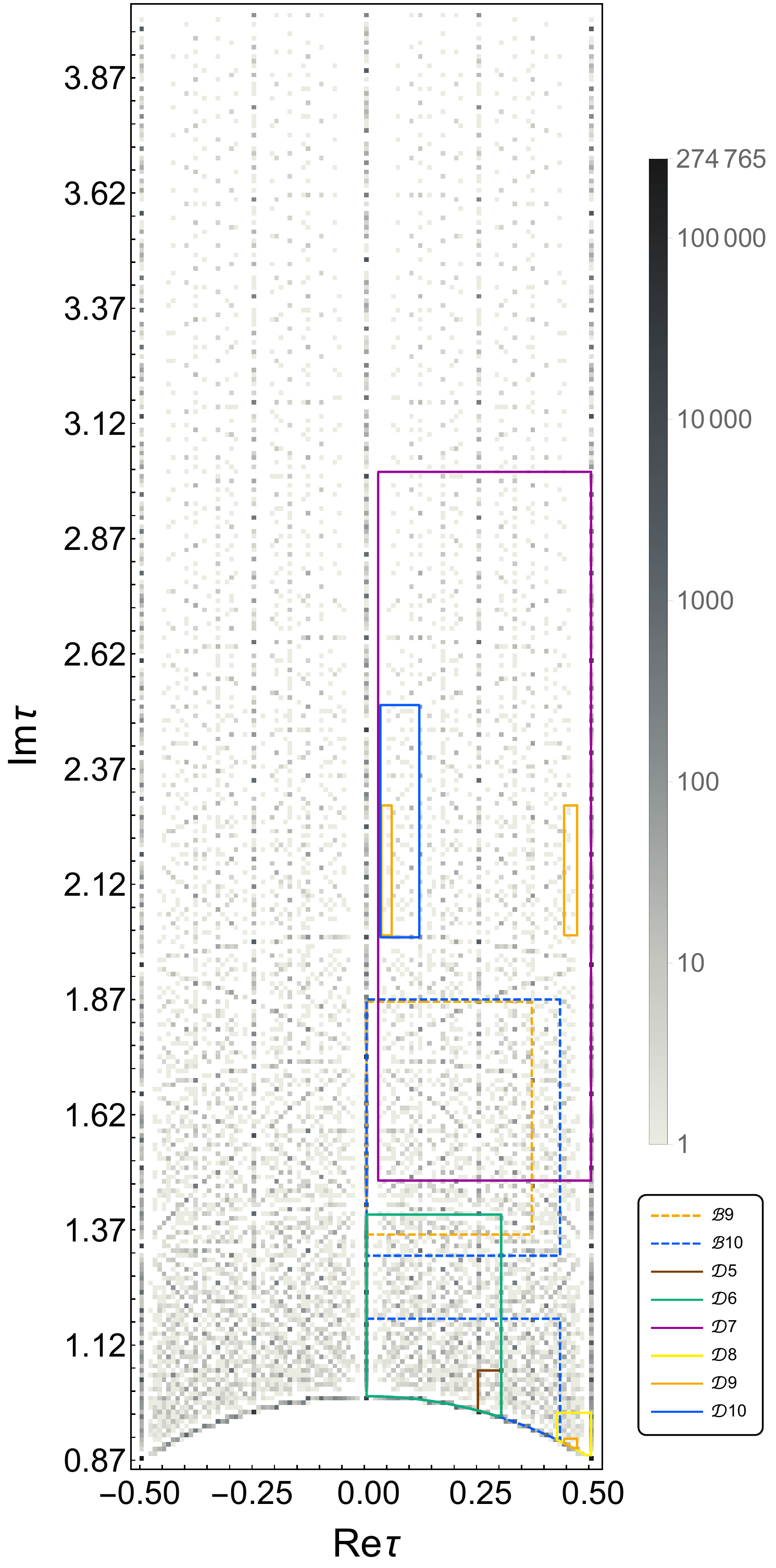}
  \end{center}
 \end{minipage}
 \begin{minipage}{0.50\hsize}
  \begin{center}
   \includegraphics[width=51.5mm]{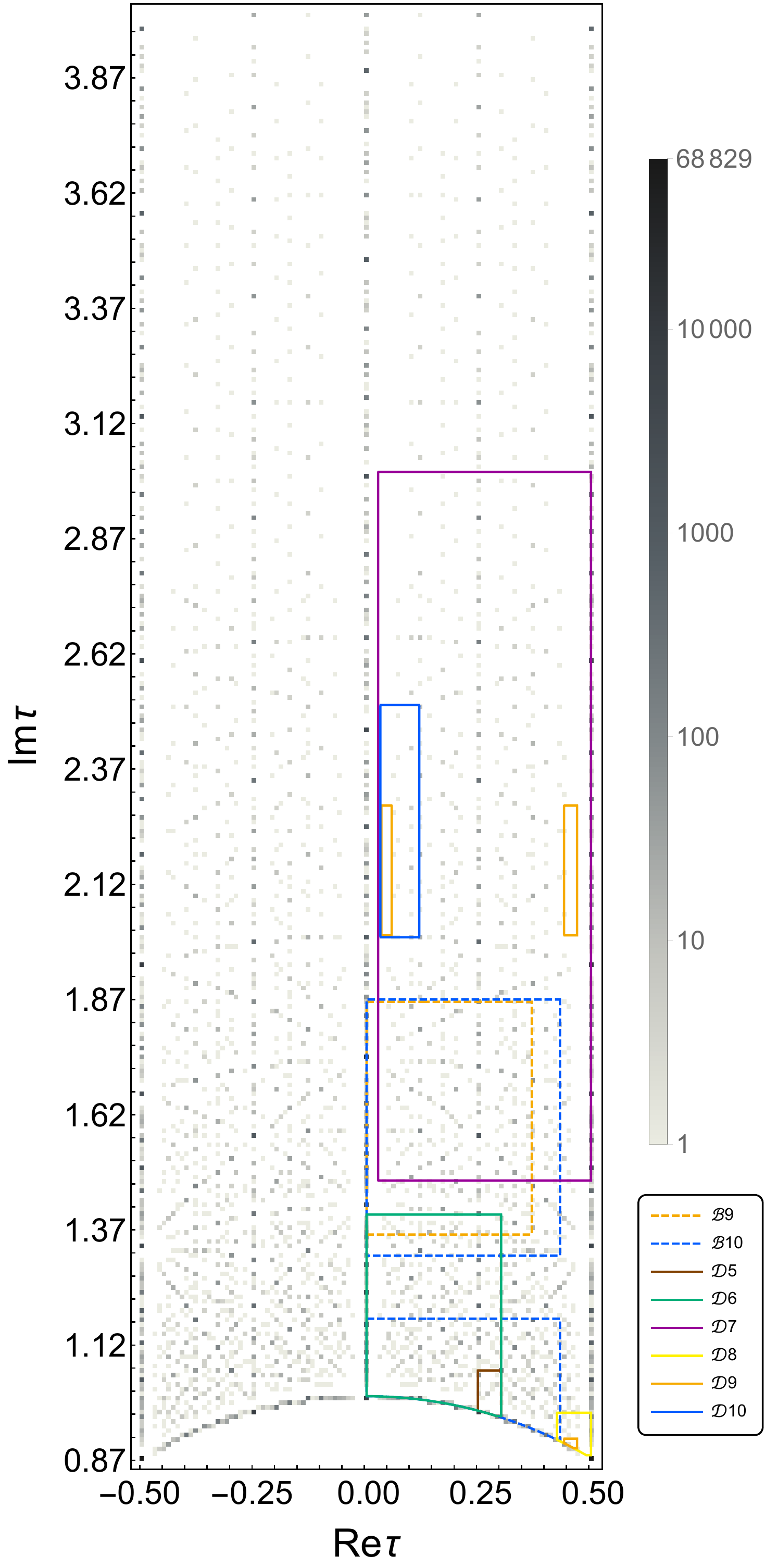}
  \end{center}
 \end{minipage}
\caption{Phenomenologically predicted moduli values vs. the complex structure modulus from the string 
landscape
for $N_{\rm flux}^{\rm max}=10,100,500,1000$, clockwise from top left.}
\label{fig:A4landscape}
\end{figure}

In Figure \ref{fig:A4landscape}, we append these phenomenologically predicted moduli values to the distributions of the complex structure modulus discussed in Section \ref{subsec:3_1}.\footnote{In Figure \ref{fig:A4landscape}, we would not consider the phenomenologically allowed region which is outside the fundamental domain and the unit circle so that the result can reflect the search method in Ref. \cite{Ding:2019zxk}.} 
Since the string landscape predicts discrete vacuum expectation values of the modulus field at the finite number of vacua, the phenomenologically allowed region is tightly constrained in the string landscape. 
Indeed, in the small $N_{\rm flux}^{\rm max}=10$ case, 
the boarder of the phenomenologically acceptable region is only allowed for ${\cal B}_9$, ${\cal B}_{10}$, ${\cal D}_{5,6,7,8}$ models, and the other regions are rejected for all models. 
When we increase $N_{\rm flux}^{\rm max}$, other phenomenologically acceptable regions are allowed.

Taking into account the results of Figure \ref{fig:A4landscape}, the vacuum expectation values of the modulus field with the highest probability in the 
string landscape for 8 classes of modular $A_4$ models are shown in Table \ref{tab:favorable_Nflux10} 
for $N_{\rm flux}^{\rm max}=10$ and Table \ref{tab:favorable_Nflux1000} 
for $N_{\rm flux}^{\rm max}=1000$, respectively. 
These moduli vacuum expectation values correspond to the theoretically favored values in the string landscape.
The probabilities of these favored moduli vacuum expectation values in (i) each model and (ii) whole fundamental domain are also displayed in Tables \ref{tab:favorable_Nflux10} and \ref{tab:favorable_Nflux1000}, 
respectively. 
Furthermore, (iii) the probability of the phenomenologically allowed region for $({\rm Re}\,\tau,{\rm Im}\,\tau)$ inside the 
string landscape is also shown in Tables \ref{tab:favorable_Nflux10} and \ref{tab:favorable_Nflux1000}.
These tables indicate that the realistic modulus values $\tau$ in ${\cal B}_9, {\cal B}_{10}, {\cal D}_{5,6,7}$ models are likely to be realized in the string landscape, in comparison with ${\cal D}_{8,9,10}$ models. 

We remark about our analysis in the modular $A_4$ models. 
First, we set the size of each bin as $(0.01, 0.01)$ in Figure \ref{fig:A4landscape} just for visibility, 
but in Tables \ref{tab:favorable_Nflux10} and \ref{tab:favorable_Nflux1000}, we show moduli vacuum expectation values up to three decimal places in the calculation of various probabilities to compare with the results of Ref. \cite{Ding:2019zxk}.
Next, there are some strongly favored points on the unit circle as seen from Tables \ref{tab:favorable_Nflux10}, \ref{tab:favorable_Nflux1000}. 
However, we excluded such points to compare with the results of \cite{Ding:2019zxk}, but in general if an allowed region contains such point on the unit circle $|\tau|=1$, it would become highly favored region in the string landscape. 
Indeed, the allowed regions of ${\cal D}_8$, (${\cal B}_{10},{\cal D}_6$) and (${\cal B}_{10},{\cal D}_5$) models allowing values around $|\tau|=1$ are favored around these region as mentioned before. 
In particular, the ${\cal D}_8$ model predicts observed values close to the $\mathbb{Z}_3$ fixed point. 
(See for the phenomenological studies around these points, Refs. \cite{Novichkov:2018ovf,Novichkov:2018yse,Okada:2019uoy,Gui-JunDing:2019wap}.)

From the viewpoint of the CP violation, the value ${\rm Re}\,\tau=1/4$ is important.
Thus, most of models, ${\cal B}_{9,10}$ and ${\cal D}_{5,6,7}$ include such points.
These models may be interesting from CP phenomenology.

\begin{table}[H]
\centering
\scalebox{1.0}{
\begin{tabular}{|c||c|c|c|c|} \hline
 & & \multicolumn{3}{c|}{Probabilities ($\%$)} \\ \cline{3-5}
Models & Favorable (Re$\tau$, Im$\tau$) & (i) & (ii) & (iii) \\
\hline 
\hline
${\rm Model}\, {\cal B}_9$  &  

$(0,\sqrt{3})$
&
80.000 & 7.547  & 9.434
\\ \hline
${\rm Model}\, {\cal B}_{10}$   &  

$(0,\sqrt{3})$
&
80.000 & 7.547  & 9.434
\\ \hline
${\rm Model}\, {\cal D}_5$  &  

None
&
- & - & 0.000
\\ \hline
${\rm Model}\, {\cal D}_6$  &  

$(0, \sqrt{3/2})$
&
100.000 & 1.887 & 1.887
\\ \hline
${\rm Model}\, {\cal D}_7$  &  

$(1/2,\sqrt{15}/2)$
&
80.000 & 7.547 & 9.434
\\ \hline
${\rm Model}\,{\cal D}_8$  &  

None
&
-  & - & 0.000
\\ \hline
${\rm Model}\ {\cal D}_9$  &  

None
&
-  & - & 0.000
\\ \hline
${\rm Model}\, {\cal D}_{10}$  &  

None
&
-  & - & 0.000
\\ \hline
\end{tabular}
}
\caption{The theoretically 
favored vacuum expectation values of $({\rm Re}\,\tau, {\rm Im}\,\tau)$ within each region for the $N^{\rm max}_{\rm flux} = 10 $ case. 
We also show the probabilities of (i) the favorable value in each model, (ii) the favorable 
value in whole fundamental domain of $SL(2,\mathbb{Z})$ moduli space and (iii) the allowed moduli region in the string landscape.} 
\label{tab:favorable_Nflux10}
\end{table}

\begin{table}[H]
\centering
\scalebox{1.0}{
\begin{tabular}{|c||c|c|c|c|} \hline
 & & \multicolumn{3}{c|}{Probabilities ($\%$)} \\ \cline{3-5}
Models & Favorable (Re$\tau$, Im$\tau$) & (i) & (ii) & (iii) \\
\hline 
\hline
${\rm Model}\, {\cal B}_9$  &  

$(0,\sqrt{3})$
&
64.560 & 7.556 & 11.704
\\ \hline
${\rm Model}\, {\cal B}_{10}$   &  

$(0,\sqrt{3})$
&
54.506 & 7.556 & 13.863
\\ \hline
${\rm Model}\, {\cal D}_5$  &  

$(3/10,\sqrt{111}/10)$
&
58.133 & 0.0886 & 0.152
\\ \hline
${\rm Model}\, {\cal D}_6$  &  

$(0,\sqrt{3/2})$
&
4.927 & 1.885 & 4.495
\\ \hline
${\rm Model}\, {\cal D}_7$  &  

$(1/2,\sqrt{15}/2)$
&
38.170 & 4.850 & 12.706
\\ \hline
${\rm Model}\,{\cal D}_8$  &  

$(1/2,\sqrt{11/12})$
&
56.216 & 0.0597 & 0.106
\\ \hline
${\rm Model}\ {\cal D}_9$  &  

$(1/22, \sqrt{2331}/22)$
&
4.762  & 0.000 & 0.003
\\ \hline
${\rm Model}\, {\cal D}_{10}$  &  

$(1/10, \sqrt{399}/10)$
&
16.466 & 0.006 & 0.037
\\ \hline
\end{tabular}
}
\caption{The theoretically 
favored vacuum expectation values of $({\rm Re}\,\tau, {\rm Im}\,\tau)$ within each region for the $N^{\rm max}_{\rm flux} = 1000$ case. 
We also show the probabilities of (i) the favorable value in each model, (ii) the favorable 
value in whole fundamental domain of $SL(2,\mathbb{Z})$ moduli space and (iii) the allowed moduli region in the string landscape.} 
\label{tab:favorable_Nflux1000}
\end{table}

\clearpage

\section{Conclusion}
\label{sec:con}

We have studied the moduli stabilization by flux compactifications on 
$T^6/(\mathbb{Z}_2\times \mathbb{Z}_2^\prime)$ orientifold 
from the viewpoint of modular flavor symmetries.
We have systematically analyzed stabilized moduli values in possible configurations 
of flux compactifications, investigating probabilities of moduli values and showing 
which moduli values are favorable from our moduli stabilization.
Then, we have 
examined their implications on modular symmetric flavor models.
%

In the bottom-up approach, the modular and CP symmetries lead to the 
construction of phenomenologically attractive models. 
Since Yukawa couplings as well as the higher-order couplings are controlled 
by the modular form of finite discrete subgroups of $SL(2,\mathbb{Z})$, 
the stabilization of complex structure moduli appearing in the Yukawa couplings 
is of particular importance to compare the predictions of modular flavor models 
with the observational data. 
These moduli values determine not only the flavor structure of quarks and leptons 
but also the CP-violating phases through the Yukawa couplings written by the modular forms. 
So far, the random search in the fundamental domain of $SL(2,\mathbb{Z})$ have 
often been carried out in the bottom-up approach. 
Our approach is to predict these moduli values and their probability distributions 
in the string landscape.

It turns out that values of the complex structure modulus $\tau$ are clustered at the $\mathbb{Z}_3$ 
fixed point $\tau=\omega$ in the fundamental region of $SL(2,\mathbb{Z})_\tau$ associated with 
the geometrical symmetry of the torus. 
Especially, the $\mathbb{Z}_3$ fixed point  is statistically favored in the weak 
string coupling and small flux regime, although the distribution of the complex 
structure modulus is correlated with that of the axio-dilaton $S$. 
It is possible to realize the other $\mathbb{Z}_2$ fixed point by focusing on 
specific vacuum expectation values of the axio-dilaton such as the $\mathbb{Z}_2$ 
fixed point in the fundamental region of the $SL(2,\mathbb{Z})_S$ moduli space 
associated with the axio-dilaton. 
In general, the modulus values with definite probabilities are intersecting points between 
$|\tau|^2=k/2$ and ${\rm Re}\,\tau=0, \pm 1/4, \pm 1/2$.

From both the top-down and bottom-up approaches, it is quite interesting that 
the $\mathbb{Z}_3$ fixed point $\tau=\omega$ as well as the $\mathbb{Z}_2$ fixed point is favored.
For example, when $\tau=\omega$, the four fixed points on the $T^2/\mathbb{Z}_2$ orbifold are equally spaced like a 
tetorahedron.
That is quite symmetric.
That affects the couplings among 
localized modes on fixed points, e.g. twisted strings in heterotic orbifold models, 
leading to the same strengths of couplings among different modes.
That is, remnant of such a geometrical symmetry appears in string-derived effective field theory.
Phenomenologically the residual $\mathbb{Z}_3$ and $\mathbb{Z}_2$ symmetries 
can appear in mass matrices of the bottom-up approach model building, constraining their forms.
In this respect, our string landscape supports analyses in modular flavor models near 
these fixed points.
In addition, it is interesting to extend such phenomenological analyses 
to other intersecting points between 
$|\tau|^2=k/2$ and ${\rm Re}\,\tau=0, \pm 1/4, \pm 1/2$.

It is found that CP-breaking stable vacua 
are statistically disfavored in the complex structure modulus sector, 
meaning that the complex structure modulus is  mostly stabilized 
at ${\rm Re}\,\tau=0$ or the boundary of the $SL(2,\mathbb{Z})_\tau$ moduli space. 
Among CP-breaking vacua, the values ${\rm Re}\,\tau=\pm 1/4$ are most favorable.
Although the  probability of CP-breaking vacua is small compared e.g., with the $\mathbb{Z}_3$ fixed point $\tau=\omega$, 
its probability increases for certain values of the axio-dilaton, e.g., ${\rm Re}\,S=1/4$.
That means a strong correlation between CP phases due to ${\rm Re}\,\tau$ and ${\rm Re}\,S$.
That suggests the possibility that 
different CP phases in the 4D effective field theory, 
 e.g., CP phases in fermion mass matrices and the strong CP phase, 
may be related with each other in superstring theory.

As for the phenomenological implication, we examine the modular $A_4$ models in 
Ref. \cite{Ding:2019zxk}, where the lepton sector has the irreducible representations of $A_4$ 
realized as a principle congruence subgroup of $SL(2,\mathbb{Z})_\tau$. 
For the 8  phenomenologically attractive models with the normal 
ordering neutrino mass spectra, we compare these phenomenological predictions 
with the distribution of the complex structure modulus. 
We find that these phenomenological results are tightly constrained 
in our results of the string landscape in which the complex structure 
modulus is distributed at the discrete vacua as shown in Fig. \ref{fig:A4landscape}. 
In this paper, we focused on modular $A_4$ models but our results in Section \ref{sec:3} 
are applicable to other modular flavor models. 
It is interesting to apply our method to other modular flavor models including 
concrete flavor models in string-derived effective field theories\footnote{See for concrete models in string-derived effective field theories, e.g., Refs. \cite{Abe:2012fj,Abe:2014vza,Abe:2015mua,Fujimoto:2016zjs,Kobayashi:2016qag}.
For example in Ref.~\cite{Abe:2014vza}, the modulus values $\tau=1.7i\simeq \sqrt{3}i$ and 
$2i$ are assumed to obtain realistic results in quark and charged lepton masses and quark mixing angles.}.

\subsection*{Acknowledgements}

T. K. was supported in part by MEXT KAKENHI Grant Number JP19H04605. H. O. was
supported in part by JSPS KAKENHI Grant Numbers JP19J00664 and JP20K14477.

\appendix
\section{Details of the numerical search}
\label{app}

In this section, we show how one can generate physically-distinct solutions to the supersymmetric conditions (\ref{eq:SUSY-min}). 
Recalling that in flux compactifications of string theory, moduli fields and fluxes are transformed under 
the modular transformations, 
flux vacua are characterized by sets of fluxes and vacuum expectation values (VEVs) of the moduli fields, 
rather than just moduli VEVs.  
Hence, we call two vacua the physically-distinct vacua when they are not related with the $SL(2, \mathbb{Z})_{S, \tau}$ transformations,  namely, 
\begin{align}
     ^\exists {\cal A} \in SL(2, \mathbb{Z})_{S} \times SL(2, \mathbb{Z})_{\tau} \text{ s.t. }  ({\rm fluxes', VEVs'}) = ({\cal A} ({\rm fluxes}), {\cal A} ({\rm VEVs})) \nonumber \\  \Leftrightarrow ({\rm fluxes, VEVs}) \sim ({\rm fluxes', VEVs'}).
\end{align}
The concrete expressions of flux transformations are given in Section \ref{sec:2}. 
Moreover, there is a degree of freedom to change overall sign of fluxes. 
To simplify the supersymmetric minimum conditions taking into account the above perspective, 
let us redefine RR and NS fluxes $\{a^0,a,b,b_0, c^0, c, d, d_0\}$ into the following 
integer variables $\{m, l, n, u, v, s, r\}$ as provided in Section \ref{sec:3}\footnote{Note that there remain only seven variables from eight fluxes and two moduli, 
since we imposed three equations (\ref{eq:SUSY-min}) on those.}. The relation between two sets is expressed as :
\begin{alignat}{4}
    r l &=  a^0,~ &  r m + s l &=  -3 a,~ &  r n + s m  &= -3 b,~ &  s n &=  -b_0, \nonumber \\ 
    u l &= c^0,~ &  u m + v l &= -3 c,~  &  u n + v m &= -3 d,~ &  v n &= -d_0.
\end{alignat}

In fact, the relation is not a one-to-one correspondence and this is problematic for counting vacua correctly. 
One need to exclude sets $\{m, l, n, u, v, s, r\}$ which gives same $\{a^0, a, b, b_0, c^0, c, d, d_0\}$ carefully as discussed later.

The D3-brane charge is now represented as
\begin{align}
    (m^2 - 4 l n) ( r v - s u ) = - 3 N_{\rm flux}. \label{eq:mlnD3tadpole}
\end{align}

As a result of the flux quantization, this implies that $N_{\rm flux}$ is a multiple of 64. 
Furthermore, the left-hand side becomes a multiple of $9 \times 64$ \cite{Kachru:2002he}, and then $N_{\rm flux}$ becomes a multiple of 192, which is not clear at this time. 
These relatively-large D3-brane charge would cause the backreactions to the geometry. 
However, we do not consider these backreactions in this paper. 

Then, we present how one can select only physically-distinct vacua in detail. At first, the flux transformations in Section \ref{sec:2} are rewritten by seven variables $\{m, l, n, u, v, s, r\}$;
\begin{itemize}
    \item $T^q \in SL(2, \mathbb{Z}): \tau \rightarrow \tau + q$
    \begin{alignat}{2}
        m &\rightarrow m -2 q l,~ & s &\rightarrow s - q r, \nonumber\\
        v &\rightarrow v - q u,~ &n &\rightarrow n - q m + q^2 l,
    \end{alignat}
    \item $S \in SL(2, \mathbb{Z})_\tau:  \tau \rightarrow - \frac{1}{\tau}$
    \begin{alignat}{4}
        m &\rightarrow -m,~ & l &\rightarrow n,~ & n &\rightarrow l,&  \nonumber \\
        u &\rightarrow -v,~& v &\rightarrow u,~ & s &\rightarrow r,~ & r &\rightarrow -s, 
    \end{alignat}
    \item $T^q \in SL(2, \mathbb{Z}): S \rightarrow S + q$ 
    \begin{alignat}{2}
        s &\rightarrow s + q v,~ & r &\rightarrow r + q u,
    \end{alignat}
    \item $S \in SL(2, \mathbb{Z})_S: S \rightarrow - \frac{1}{S}$
    \begin{alignat}{4}
        u &\rightarrow -r,~ & v &\rightarrow -s,~ & s &\rightarrow v,~ & r&\rightarrow u,
    \end{alignat}
\end{itemize}
where $q \in \mathbb{Z}$ and invariant elements are simply omitted.
From these, we can fix degrees of freedom of $T$-transformations. 
\begin{itemize}
    \item $T^q \in SL(2, \mathbb{Z})_\tau$\\
    As discussed in Section \ref{sec:2}, we can focus on only the $l \neq 0$ case. Hence, there is the equivalence relation:
   \begin{align}
       m \sim m - 2 q l,
   \end{align}
    i.e., there are only $2l$ independent choice of $m$, and we can freely set $m$ as 
    \begin{align}
        m = -l, -l+1, \dots, l-1.
    \end{align}
    
    \item $T^q \in SL(2, \mathbb{Z})_S$\\
    In this case, $v=0$ is possible but $u=0=v$ leads unstabilized $S$. Hence we have to consider these two cases:
    \begin{align}
        s &\sim s + q v ~ (v \neq 0),\\
        r &\sim r + q u ~ (v = 0).
    \end{align}
    
\end{itemize}

From now, we focus on the $l, n > 0$ case. The $l, n < 0$ case is completely analogous, so that we simply omit it. 
However, we included both results in the actual calculation. 
Recall that, since we require $\tau$ given in Eq. (\ref{eq:lnpositive}) to be complex, $m^2 - 4 l n$ must be negative 
in the $l, n > 0$ case. 
Furthermore, we can impose more severer condition on it:
\begin{align}
    m^2 - 4 l n < - 3 l^2.
\end{align}
Since the inside of the unit circle and its outside are connected by $S$-transformations and $T$- transformations 
have been already fixed, it is enough to search for ${\rm Im}\,\tau \geq \frac{\sqrt{3}}{2}$ region.

Let us come back to Eq. (\ref{eq:mlnD3tadpole}). 
Since $r v- s u$ must be integer, 
\begin{align}
    \left| m^2 - 4 l n \right| \leq 3 N_{\rm flux} 
\end{align}
holds. 
Taking into consideration that $N_{\rm flux} \geq 0$ holds for the supersymmetric solutions, we conclude
\begin{align}
    l &= 1, \dots, \sqrt{N_{\rm flux}},  \\
    \frac{3 l^2 + m^2}{4l} &\leq n \leq \frac{3 N_{\rm flux} + m^2}{4l}. \label{eq:condition of n}
\end{align}
To constrain other fluxes $\{r, s, u, v\}$, we look at the expression of $S$ in Eq. (\ref{eq:Svev}). As with the $\tau$ case, it is enough to require ${\rm Im}S \geq \frac{\sqrt{3}}{2}$, thereby
\begin{align}
   \frac{3N_{\rm flux}}{m^2 - 4ln} = s u - r v \leq - \frac{\sqrt{3}}{2} \frac{\left| u \tau + v \right|^2}{{\rm Im} \tau}.
\end{align}
By expanding $\left| u \tau + v \right|^2 = \left(u{\rm Re}\,\tau + v\right)^2 + u^2\left({\rm Im}\,\tau \right)^2$ and reducing the above condition for $u, v$, we find that
\begin{align}
    \left|u\right| &\leq \sqrt{\frac{ \sqrt{3} N_{\rm flux} }{ 2 l^2 ({\rm Im}\,\tau)^3} }, \label{eq:condition of u}  \\
    - \sqrt{\frac{ \sqrt{3} N_{\rm flux} }{ 2 l^2 {\rm Im}\,\tau } - (u{\rm Im}\,\tau)^2 } - u{\rm Re}\,\tau &\leq v \leq  \sqrt{\frac{ \sqrt{3} N_{\rm flux} }{ 2 l^2 {\rm Im}\,\tau } - (u{\rm Im}\,\tau)^2 } - u{\rm Re}\,\tau,
    \label{eq:condition of v}
\end{align}
with ${\rm Re}\,\tau = -\frac{m}{2l}$. As discussed before, we have to divide in two cases, namely (i) $v \neq 0$ and (ii) $v=0$.
\begin{itemize}
    \item (i) $v \neq 0$ case\\
        In this case, we adopt $s \sim s + q v$ so that
        \begin{align}
            s = 1, 2, \dots, |v|-1.
        \end{align}
        Then one can obtain $r$  from the expression of $N_{\rm flux}$ (\ref{eq:mlnD3tadpole}) as 
        \begin{align}
            r = \frac{1}{v} \left( s u - \frac{3 N_{\rm flux}}{m^2 -4 l n}\right). \label{eq:expression of r}
        \end{align}
        
    \item (ii) $v = 0$ case\\
        Similarly, one can obtain $s$ from $N_{\rm flux}$ as 
        \begin{align}
            s = \frac{3 N_{\rm flux}}{(m^2 -4 l n ) u}, 
            \label{eq:expression of s}
        \end{align}
        and we adopt $r \sim r + q u$ so that 
        \begin{align}
            r = 1, 2, \dots, |u| - 1.
        \end{align}
\end{itemize}
Note that it is not clear that whether $r, s$ are integer or not at this stage, therefore one has to check the integrability  constraints in the actual calculation.

Lastly, we summarize the protocol to obtain all the physically-distinct solutions:
\begin{enumerate}
    \item Set $k \in 1, \dots, 1000$ so that $N_{\rm flux} = 192 \times k$.
    \item Find a list of divisors of $3N_{\rm flux}$ (${\rm div} 3N_{\rm flux}$) and think its element as $m^2 - 4 l n$. Choose elements of $g \in {\rm div} 3 N_{\rm flux}$ satisfying $g \equiv 0, 1\pmod 4$ to identify them with $m^2 - 4 l n$.
    \item Set $l = 1, 2, \dots, \lfloor \sqrt{N_{\rm flux}} \rfloor$.
    \item For each $l$, set one $m$ from $-l, -l+1, \dots, l-1$.
    \item Set $n = \frac{{\rm div} 3 N_{\rm flux} - m^2}{-4l}$ and check whether $n$ satisfies Eq. (\ref{eq:condition of n}). If not, return to 3 and repeat 4. Then, $\tau$ is determined.
    \item Since $u$ must reside in the range (\ref{eq:condition of u}) and $u l \in 8\,\mathbb{Z}$, it is enough to consider $u$ as an element in the following range 
    \begin{align}
        - \left\lfloor \frac{\frac{\sqrt{3}N_{\rm flux}}{2l^2 ({\rm Im}\,\tau)^3}}{\operatorname{lcm} (l, 8)} \times l\right\rfloor \frac{\operatorname{lcm}(l, 8)}{l} \leq u \leq  \left\lfloor \frac{\frac{\sqrt{3}N_{\rm flux}}{2l^2 ({\rm Im}\,\tau)^3}}{\operatorname{lcm} (l, 8)} \times l \right\rfloor \frac{\operatorname{lcm}(l, 8)}{l},
    \end{align}
    whose step for increment is $\frac{\operatorname{lcm}(l, 8)}{l}$.  
    \item Let us rewrite the range for $v$ (\ref{eq:condition of v}) as $ m_v \leq v \leq M_v$ for simplicity. Taking into account the requirement $v n \in 8\,\mathbb{Z}$, it is enough to consider $v$ as an element in the following range
    \begin{align}
        \left \lceil \frac{ m_v  n}{\operatorname{lcm}(n,8)} \right \rceil \times \frac{\operatorname{lcm}(n,8)}{n} \leq v \leq \left \lfloor\frac{M_v n}{\operatorname{lcm}(n,8)} \right \rfloor \times \frac{\operatorname{lcm}(n,8)}{n}
    \end{align}
    whose increment step is now given by $\frac{\operatorname{lcm}(n,8)}{8}$, and $v$ becomes an element in the list constructed by the above bounds and increments. Here one has to impose all the constraints: $ u\neq 0 \lor v\neq 0,u m + v l\in 24\,\mathbb{Z}, u n + v m \in 24\,\mathbb{Z}.$  
    \item For $v\neq 0$ case, one can set $s$ as an element of $0, \dots, |v|-1$. For each configuration ($m, l, n, u, v, s$), determine $r$ by (\ref{eq:expression of r}) in the end. Here one has to impose all the constraints: $r \in \mathbb{Z}, rl \in 8\,\mathbb{Z}, r m + s l \in 24\,\mathbb{Z}, r n + s m \in 24\,\mathbb{Z}.$  
    
    For $v = 0$ (and $u \neq 0$) case,  one can set $r$ as an element of $0, \dots, |u|-1$. For each configuration ($m, l, n, u, v, r$), determine $s$ by (\ref{eq:expression of s}) in the end. Here one has to impose all the constraints: $s \in \mathbb{Z}, sn \in 8\,\mathbb{Z}, r m + s l \in 24\,\mathbb{Z}, r n + s m \in 24\,\mathbb{Z}.$  
    \item At last, one has a set of whole $T$-independent fluxes so that one can calculate the VEVs given as functions of fluxes. Then one can bring them to $-\frac{1}{2} \leq \text{Real part} < \frac{1}{2}$ in terms of the corresponding $T$-transformation.
    Although this leads to the rectangular region of moduli space, one can employ degrees of freedom of $S$-transformations to cut off inside the unit circle and reduce the number of solutions on the unit circle except for two fixed points by half (or one can cut off solutions on the right side of it, but the former is suitable to draw symmetric figures). 
    For each fixed point, one should take account of enhanced symmetries. Hence one has to reduce the number of solution at $Z_2$ and $Z_3$ fixed points to one-half and one-third, respectively\footnote{One can see that $(T^{q = -1}S)^2_{SL(2, \mathbb{Z})_\tau} = (T^{-1}S)_{SL(2, \mathbb{Z})_S}$ holds at $(\tau, S) = (\mathbb{Z}_3,\mathbb{Z}_3)$ up to overall sign of fluxes. Hence the degeneracy is three, not nine. Similarly, the degeneracy on $(\mathbb{Z}_2,\mathbb{Z}_2)$ is two, not four.}.
    Finally, one need to truncate flux configurations $\{m, l, n, u, v, s, r\}$ which degenerate into same $\{a^0, a, b, b_0, c^0, c, d, d_0\}$ and ignore overall sign of them. Then, we obtain whole the physically-distinct solutions.
\end{enumerate}

\end{document}